# Metal-Insulator Transitions and non-Fermi Liquid Behaviors in 5d Perovskite Iridates

Abhijit Biswas[1], Ki-Seok Kim[1,2], and Yoon Hee Jeong[1,a]

[1]Department of Physics, POSTECH, Pohang, 790-784, South Korea
[2]Institute of Edge of Theoretical Science, POSTECH, Pohang, 790-784, South Korea
[a]Corresponding Author:yhj@postech.ac.kr

**Abstract**

Transition metal oxides, in particular, 3d or 4d perovskites have provided diverse emergent physics that originates from the coupling of various degrees of freedom such as spin, lattice, charge, orbital, and also disorder. 5d perovskites form a distinct class because they have strong spin-orbit coupling that introduces to the system an additional energy scale that is comparable to bandwidth and Coulomb correlation. Consequent new physics includes novel $J_{eff}$ = 1/2 Mott insulators, metal-insulator transitions, spin liquids, and topological insulators. After highlighting some of the phenomena appearing in Ruddlesden-Popper iridate series $Sr_{n+1}Ir_nO_{3n+1}$ ($n$ = 1, 2, and ∞), we focus on the transport properties of perovskite $SrIrO_3$. Using epitaxial thin films on various substrates, we demonstrate that metal-insulator transitions can be induced in perovskite $SrIrO_3$ by reducing its thickness or by imposing compressive strain. The metal-insulator transition driven by thickness reduction is due to disorder, but the metal-insulator transition driven by compressive strain is accompanied by peculiar non-Fermi liquid behaviors, possibly due to the delicate interplay between correlation, disorder, and spin-orbit coupling. We examine various theoretical frameworks to understand the non-Fermi liquid physics and metal-insulator transition that occurs in $SrIrO_3$ and offer the Mott-Anderson-Griffiths scenario as a possible solution.

# 1. Introduction

Transition metal oxides (TMOs), ranging from simple binary oxides to more complex ternary or quaternary compounds, have been a subject of intense activities in condensed matter physics and materials science [1-5]. Of various transition metal oxides, *perovskites* and their variations are a particularly important class because they display a rich spectrum of various competing phases and physical properties. Structurally, in perovskites with the chemical formula $ABO_3$, large rare-earth or alkali-metal cations occupy the A-sites (tetrahedral oxygen interstices) and transition metal cations occupy the B-sites (octahedral oxygen interstices). Each transition metal ion is thus surrounded by six oxygen anions to form a $BO_6$ octahedron, and the octahedra are connected to each other three-dimensionally. This transition metal–oxygen network mostly determines the electronic properties; while the close-packed A cations and oxygen anions provide structural stability. One characteristic feature of the perovskite family is that numerous transition metal ions of different size and valence are allowed in the same structure. Along with this elemental diversity, mutually-interacting quantum degrees of freedom such as lattice, spin, charge, and orbital lead to numerous emergent physical properties in perovskites. For example, the conducting properties of perovskites, range from insulating to semiconducting to metallic and finally to superconducting. The electrical properties of perovskites are equally diverse; they also display various magnetic and multifunctional properties [6-14]. These unparalleled varieties in the electronic properties of perovskites hold tremendous application potential and thus provide a continuous impetus for research in the field.

The network of $BO_6$ octahedra in transition metal perovskites strongly influences their electronic properties because *d*-orbitals form bands with characteristic effective masses and different strengths of on-site Coulomb interactions. Indeed, diverse phenomena have been treated in terms of various relevant energy scales such as bandwidth ($W$) and Coulomb repulsion ($U$), and the relevant efforts can be termed *d*-orbital physics [15,16]. Disorder ($D$) which is unavoidable in real solids also has an important influence on the physics of transition metal oxides.

In materials science, numerous studies of 3$d$- and 4$d$-based transition metal perovskites have been conducted to find ways to exploit their functionalities that depend sensitively on structural distortion and crystal chemistry. However, perovskite studies have been limited mainly to those that contain 3$d$ or 4$d$ transition metal elements. In this regard, 5$d$ perovskites form a special class because the strong spin-orbit coupling (SOC, $\Lambda$) introduces to the system an additional energy scale that is comparable to bandwidth and Coulomb correlation [17]. For example, metal oxide compounds containing 5$d$ iridium (e.g. $Sr_2IrO_4$, $Sr_3Ir_2O_7$, $SrIrO_3$, $BaIrO_3$, $Na_2IrO_3$, $Eu_2Ir_2O_7$, $Na_4Ir_3O_8$, and $Sr_2GdIrO_6$) display a variety of emerging properties (Table 1). Of the various exotic properties of 5$d$ perovskites, the transport properties and, in particular, the metal-insulator transitions (MIT) in iridate perovskites are the subject of the present chapter.

**TABLE 1.** 5$d$ transition metal iridium (Ir) based oxides, which are mostly insulators with exotic magnetic states.

| Structure | Compound | Properties |
|---|---|---|
| Layered | $Sr_2IrO_4$ | $J_{eff}$ = 1/2 Mott insulator, |
| | $Sr_3Ir_2O_7$ | complex magnetism, |
| | $Ba_2IrO_4$ | lattice magnetoresistance |
| Hexagonal | $BaIrO_3$ | $J_{eff}$ = 1/2 Mott insulator, |
| | $Ca_5Ir_3O_{12}$ | non-Fermi liquid, magnetism, |
| | $Sr_3NiIrO_6$ | geometric frustration |
| Pyrochlore | $Y_2Ir_2O_7$ | Topological insulator, |
| | $Bi_2Ir_2O_7$ | strong magnetic instability, |
| | $Eu_2Ir_2O_7$ | metal-insulator transition |
| Honeycomb | $Na_2IrO_3$ | Topological insulator, |
| | $Li_2IrO_3$ | zig-zag magnetic order |
| Orthorhombic | $SrIrO_3$ | Metal-insulator transition, |
| | $CaIrO_3$ | non-Fermi liquid, semi-metal |
| Double perovskite | $Sr_2YIrO_6$ | Correlated insulator, |
| | $Sr_2GdIrO_6$ | exotic magnetism, |
| | $La_2MgIrO_6$ | spin wave excitations |
| Kagome | $Na_4Ir_3O_8$ | Spin liquid |

Simple extrapolation of the characteristics of 3*d* or 4*d* systems fails to predict the characteristics of 5*d* systems. Moving down the periodic table from 3*d*→4*d*→5*d*, the orbitals in the solids that contain the corresponding *d*-orbitals become increasingly extended and so does the bandwidth ($W_{3d} < W_{4d} < W_{5d}$). Along with an increase in the bandwidth, the associated on-site Coulomb repulsion would decrease correspondingly ($U_{3d} > U_{4d} > U_{5d}$). From these considerations, one may expect higher metallicity and less magnetic instability for the materials with more extended 5*d* orbitals, compared with the systems containing 3*d* or 4*d* elements, because the Stoner criterion in these compounds favors paramagnetic metallic states. Surprisingly, however, the properties of the 5*d* based transition metal oxides exhibit extremely rich behaviors (Table 1). The Ruddlesden-Popper (RP) series of iridium-based $Sr_{n+1}Ir_nO_{3n+1}$ (*n* = 1, 2, and ∞) as well as other structural families such as pyrochlores, kagome-type lattices, honeycomb-type structures and double perovskites are mostly insulating (few are metallic) but may have exotic physical properties, including novel Mott insulator, lattice-driven magnetoresistance, giant magnetocaloric effect, quantum criticality, charge density wave, geometrically-frustrated magnetism, possible topological insulator, and Weyl semi-metal [17-19]. These diverse behaviors in iridates require that a new area of physics must be considered; this is SOC ($\Lambda_{3d} < \Lambda_{4d} < \Lambda_{5d}$). In iridates, the energy scales of *W*, *U*, and *Λ* are all comparable and thus compete with each other. In addition, *D* is also important, and the emergent properties are the results of strong interplay among Coulomb interaction, SOC, and disorder.

Electrical conductivity is one of the properties that used to characterize or to classify solids, so a transition from a metal to insulator or vice versa as a function of a control parameter such as temperature, pressure, or magnetic field has been a topic of interest to the condensed matter physics community for several decades [20-24]. However, despite enormous efforts, understanding of the MIT at the microscopic level is still under debate. In transition metal oxides and strongly-correlated systems, the MIT is often accompanied by a change in structural or magnetic symmetry. In contrast, in the presence of a sufficient amount of disorder, MIT is often not associated with any uniform ordering or change in symmetry. Perhaps the most important mechanisms that underlay MITs in correlated systems are interaction-driven or correlation-driven Mott localization, magnetic-order-driven Slater insulator, and disorder-driven Anderson localization. In perovskites that involve heavy 5d transition metal elements (e.g. Re, Os, Ir), SOC comes as another parameter and becomes comparable in strength to other relevant energy scales. All of these different parameters in the system compete with each other so their interplay stabilizes new exotic ground states. In this chapter our foremost aim is to provide a brief description of current research on MITs that occur in Ir-based perovskites. Our emphasis will be on a model system ($SrIrO_3$) to provide experimental evidence that the interplay among correlation, SOC, and disorder is important in achieving different MITs. We also wish to present our understanding about the observed non-Fermi liquid physics and MITs to provide insight and motivation for further activities in this rapidly developing, yet poorly understood field of strong spin-orbit-coupled 5d-based oxide physics.

## 2. Metal-insulator transitions and representative types of insulators

### 2.1 Metal-insulator transitions

Electrical resistance $R$ (or resistivity $\rho$ if intrinsic quantity is used) is a key property that characterizes materials, which are typically classified as metals or insulators. A very good metal (e.g. Cu) can have an electrical resistivity as small as $\rho$ ☐$10^{-10}$ Ω·cm, whereas a good electrical insulator (e.g. quartz) has a resistivity as high as $\rho$ ☐$10^{20}$ Ω·cm. The causes of this huge difference in resistivity are now well understood in modern solid state physics. The difference between the resistivity's of metals and those of insulators implies that MITs may be accompanied by a large change in $\rho$ (up to several orders of magnitude). However, many MITs of current interest are often accompanied not by a large change in magnitude but by a qualitative change in conducting behaviors.

Metals can be distinguished from insulators by their different responses of $R$ to temperature $T$. Metals are defined as materials in which $R$ decreases as $T$ decreases ($dR/dT > 0$), whereas insulators are materials in which $R$ inecreases as $T$ decreases ($dR/dT < 0$) (Figure 1). Strictly, the ultimate distinction between the metal and insulator can be made at absolute zero; a metal would continue be conductive whereas an insulator would lose its conductivity. MITs of fundamental interest change a system from a phase with d*R*/d*T* < 0 to another with d*R*/d*T* > 0, or vice versa.

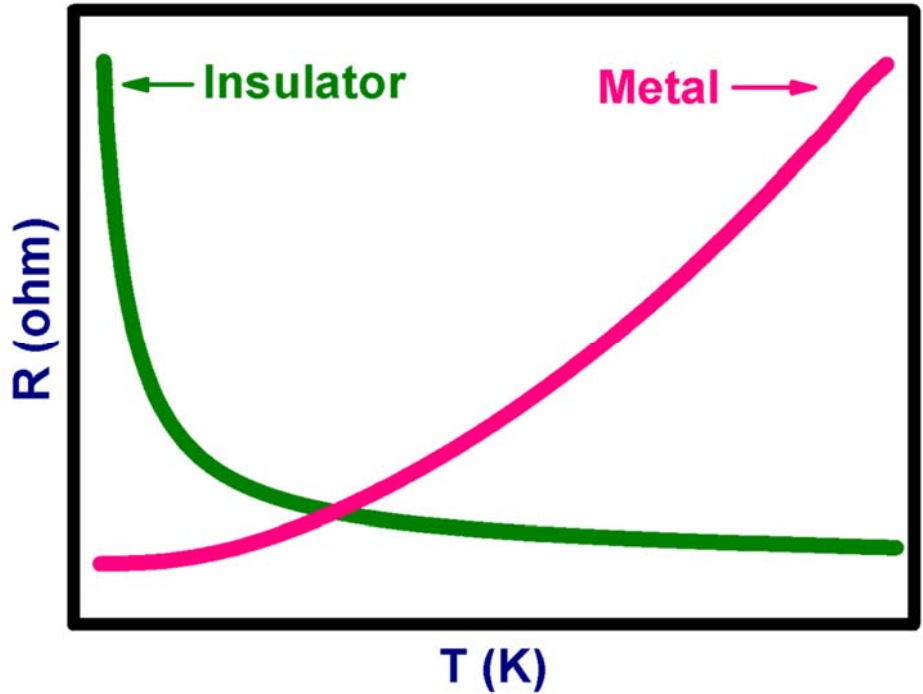

**FIGURE 1.** Metals are broadly defined as materials with d*R*/d*T* > 0, whereas insulators are as ones with d*R*/d*T* < 0. *R*: resistance; and *T*: temperature. At absolute zero, the resistance of an insulator would reach infinity with zero conductivity while a metal would still possess finite conductivity.

The conventional theory of electrical transport was first formulated by Drude [25], immediately after the discovery of the electron. The semi-classical Drude theory of electronic conductivity was built on the idea of the kinetic theory of gases, considering a metal as a gas of electrons. A key concepts in that description are the mean free path $l$ (i.e., the average length that an electron travels between successive collisions), and relaxation time $\tau$ (i.e., the average time between successive collisions). According to the Drude theory, the electronic conductivity $\sigma$ (= $1/\rho$) is directly proportional to $\tau$ as $\sigma = \frac{ne^2\tau}{m}$ where *n* is the density of electrons, *e* is the electron charge and *m* is electron mass. For good metals $l \sim 100$ nm, and $\tau \sim 10^{-14}$ sec. This semi-classical Drude model is still used today as a quick way to estimate a material's property.

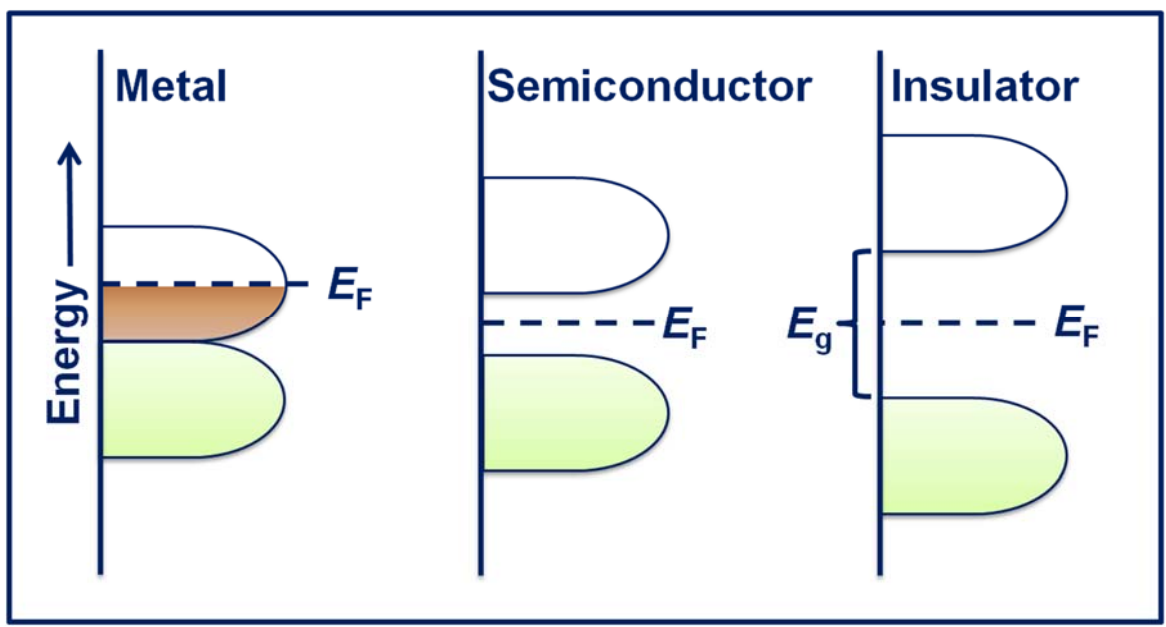

**FIGURE 2.** Schematic band diagram of metal, semiconductor and insulator. $E_F$, and $E_g$ are the Fermi energy and band-gap. A semiconductor is an insulator with a small energy gap. Upper one: conduction band (CB); lower one: valence band (VB).

Quantum mechanics has been used to clarify the transport properties of solids, and now metals, semiconductors, and insulators are classified according to the band theory of solids (Figure 2). In metals, Fermi energy $E_F$ is within the conduction band, whereas in insulators, $E_F$ separates the electronic band into an empty upper and filled lower band. For example, quartz is an insulator with a band gap $E_g \sim 8.9$ eV, whereas Cu is a metal. Si is also an insulator, but has a small $E_g \sim 1.1$ eV and thus is called a semiconductor. Although band theory successfully describes the conducting nature of many materials, it fails to account for the behaviors of many transition metal oxides that have relatively narrow separation between conduction and valence band. Many materials with partially-filled *d*-orbitals and an odd number of electrons per lattice site, which should be metals according to band theory, are actually insulators (e.g. NiO, $V_2O_3$, $Fe_3O_4$). As the band theory considers single electron, and thus may not adequate considering many-body effects such as electron-electron interaction that occur in these complex materials. In addition, *D*, which is unavoidable in real solids, must also be considered. Many theoretical explanations have been proposed to account for the metallic and insulating properties of so-called correlated materials; we briefly summarize these ideas in this section.

## 2.2 Mott insulator

In 1937, de Boer and Verwey found that many transition-metal-oxides, (e.g., NiO, MnO, FeO), show insulating features despite the fact that these oxides have partially filled *d*-bands [26]. Soon after their discovery, Peierls suggested that strong Coulomb repulsion between electrons could be the origin of the insulating behaviors [27]. He remarked that *"it is quite possible that the electrostatic interaction between the electrons prevents them from moving at all. At low temperatures the majority of the electrons are in their proper places in the ions. The minority which have happened to cross the potential barrier find therefore all the other atoms occupied, and in order to get through the lattice have to spend a long time in ions already occupied by other electrons. This needs a considerable addition of energy and so is extremely improbable at low temperatures"* [20]. Peierl's speculation aroused interest in so-called strongly-correlated systems. In 1949, Mott offered a theoretical explanation of how electron-electron correlation could yield an insulating state, now called a Mott Insulator [21].

Consider a lattice model with a single electron orbital at each site. If the electron-electron interactions are not considered, a single band would be formed from the overlap of the atomic orbitals: when two electrons, one with spin up (↑) and the other with spin down (↓), occupy each site, the band becomes full. However, when two electrons occupy the same site, they would feel a large Coulomb repulsion, which can be explained using the Hubbard Hamiltonian (Eq. (1)):

$$H = -t \sum_{\langle ij\sigma \rangle} \left( c_{i\sigma}^{\dagger} c_{j\sigma} + h.c. \right) + U \sum_{i} n_{i\uparrow} n_{i\downarrow} - \mu \sum_{i,\sigma} n_{i\sigma} \qquad (1)$$

where $c_{i,\sigma}^{\dagger}$ and $c_{j,\sigma}$ ($\sigma = \uparrow\ or\ \downarrow$) are respectively the creation and annihilation operators for electrons on the site *i* and *j* with spin $\sigma$. Electron hopping between nearest neighbor sites occurs with hopping constant −*t*. *U* is the amount of energy for each pair of electrons that occupy the same lattice site, and represents on-site Coulomb correlation. $n_{i\sigma}$ is the number operator and $\mu$ is the chemical potential.

The ratio between *t* and *U* determines whether the system is a metal or an insulator. When electron-electron correlation is negligible compared to hopping, *t/U* >> 1, the electrons tunnel between the sites without hindrance and the system is metallic. When *t/U* << 1, electron-electron correlation is strong, the electrons are localized due to Coulomb interaction and the system becomes an insulator. In the Hubbard model, at half filling $<n_{i\sigma}> = \frac{1}{2}$, the Mott-Hubbard insulating phase with one electron per site would appear.

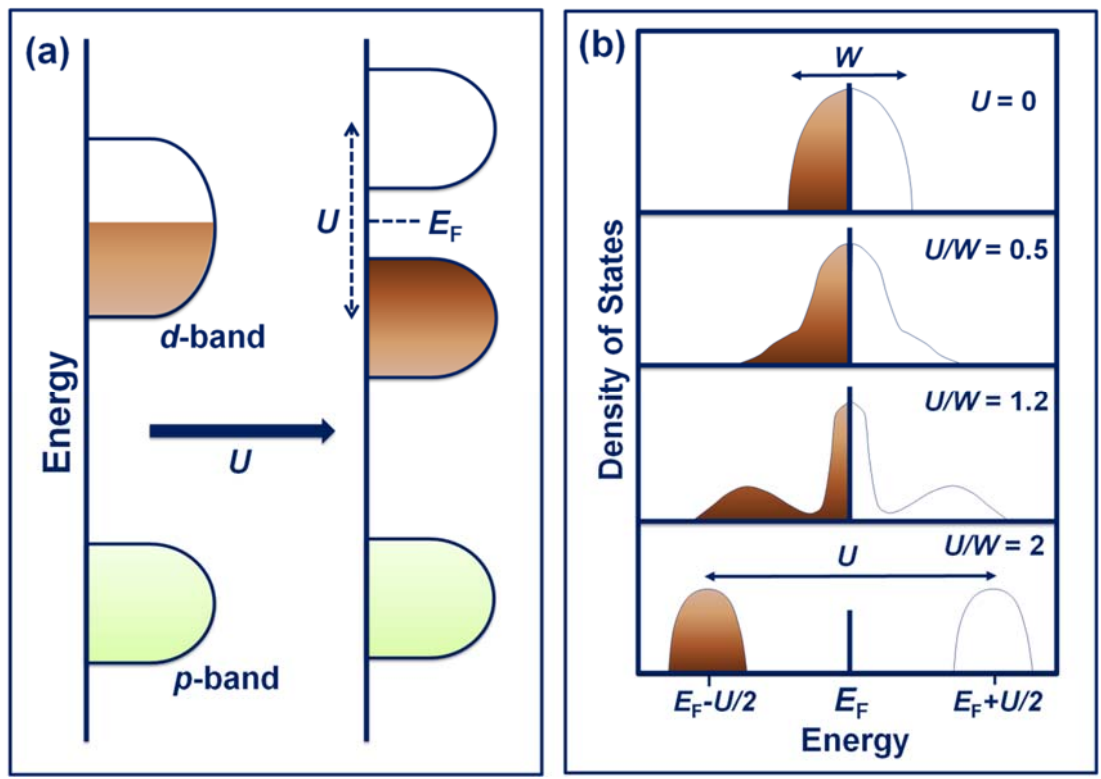

**FIGURE 3.** (a) Schematic representation of the energy levels for a Mott-Hubbard insulator where on-site Coulomb interaction *U* splits the *d*-band into lower Hubbard band and upper Hubbard band. (b) Evolution of the density of states (DOS) of electrons as a function of *U/W* (*W* = bandwidth) as the system evolves from a metal to an insulator. Adapted with permission from [28]. © 2009, AIP Publishing LLC

In transition metal oxides, often oxygen *p*-bands remain unchanged when interaction (*U*) occurs (Figure 3(a)). In contrast, the *d*-band splits into two sub-bands (upper Hubbard band and lower Hubbard band) and the half-filled state becomes an insulator with the opening of a charge gap (Figure 3(b)). The Hubbard model allows two types of MIT: the filling-control metal-insulator transition (FC-MIT), which originates from variation of electron concentration or chemical potential ($\mu/U$); and a bandwidth-control metal-insulator transition (BC-MIT), which originates from variation of hopping energy or bandwidth ($t/U$).

### 2.3 Slater insulator

Mott insulators in the previous section are electron systems with an odd number of electrons per unit cell; despite the fact that although band theory suggests that these systems would be metallic, they are insulators. Mott insulators belong to the $U \gg t$ regime in the Hubbard model. However, another aspect must be considered to explain the insulating property even in the Hubbard model. In fact, many insulating systems, especially transition metal oxides (e.g. $Na_2IrO_3$ [19]) often have antiferromagnetic ground states; Slater focused on this point. He proposed that formation of spin density waves (long-range magnetic order) due to electron-electron interaction may be an origin of the insulating phase itself [29]. The difference between a Mott insulator and a Slater insulator is that for a Mott insulator the system remains insulating even above Néel temperature $T_N$, whereas a Slater insulator the system should be metallic above $T_N$.

The principle behind a magnetic-order-driven Slater insulator may be easily understood from the band picture. Slater insulators belong to the the $t \gg U$ regime, which is opposite to the case for Mott insulators. Suppose for simplicity that the lattice is half filled and thus on average each site holds one electron. A low energy state for a half-filled system ensues if a periodicity doubling of the lattice occurs (i.e., the Brillouin zone is halved). At the new Brillouin zone boundary, an energy gap for charge excitations occurs and the system becomes insulating. According to Slater, such an insulating behavior is closely connected with the appearance of magnetic order at $T_N$. For example, electrons in a bipartite lattice [30], i.e., one that consists of two interpenetrate sub-lattices A and B in such a way that the nearest neighbors of any site are members of the opposite sub-lattice, for example, a rock salt arrangement in a simple cubic lattice. Thus, the nearest neighbors of an electron from A are those from B, and vice versa. Because ↑ and ↓ electrons mutually repel due to Coulomb interaction, they become preferentially arranged on alternating A and B sub-lattice sites. Hence spins form a spin-density wave (SDW) whose wave vector is commensurate with the lattice. Because the electrons tend to avoid each other, the potential energy increases, and this gain is balanced by a loss in kinetic energy because of localization of electrons. The spin-density wave provides a necessary periodicity-doubling potential. With an increase in temperature, thermal fluctuations affect the ordering and narrow the energy gap. Eventually at $T_N$, which is typically ~$10^2$ K, the ordering is destroyed and the insulating property disappears.

### 2.4 Anderson insulator

Defects and impurities are unavoidable in real materials, and disorder can never be neglected in reality. Thus, disorder-induced MIT is a subject of continuing interest in condensed matter physics. Anderson in 1958, initiated the field of so-called localization by arguing that sufficiently strong randomness will localize all the electronic states within a given band and that diffusion may be completely suppressed, thereby leading to an MIT at $T$ = 0 K [31]. In the absence of disorder, even a small amount of hopping can delocalize the electrons. However, in the presence of sufficient disorder, the hopping process only spreads an initially-localized state over a finite distance; defined as the localization length $\xi_0$. As a result, the states at the band tail become localized and sometimes even a whole band becomes localized. Mott argued that there must exist a critical energy $E_C$ called mobility edge, which separates the localized states from the extended ones (Figure 4(a)) [21]. If $E_F < E_C$, the system is an insulator; otherwise it is a metal.

The basic idea behind the localization phenomenon is rather straightforward. Suppose an electron propagates in a disordered medium from point A to point B. To obtain the probability that an electron reaches point B, the probabilities of all possible paths from A to B must be summed. In a disordered medium, the phases of the interference terms may be so random that on average they mutually cancel; the resultant state can be explained by the diffusion model of conductivity but, this simple argument may not be always applicable. Imagine a wave that travels from point A along a random path to point B and then goes back to A. Two possible paths are illustrated (Figure 4(b)): a randomly chosen path and the same path traversed in time reversed sense. The two paths interfere constructively and should be treated coherently as long as time reversal symmetry is preserved. Then the probability of the electron's return to A is twice as large as it would have been if probabilities were added (first squared and then added). The enhanced back-scattering, known as weak localization, reduces the conductance between A and B by increasing the electron's likelihood of returning to its starting point, and eventually leads to localization [32]. At high temperatures, coherence is lost due to thermal vibrations and back-scattering effects diminish. However, at low temperatures, thermal vibrations and inelastic scattering cease and other channels through which electrons can exchange energy become interrupted, so quantum interference immobilizes electrons to induce localization transitions. In 1979, Abrahams et al., developed the phenomenological scaling theory of localization [33, 34]. In metallic systems in two dimensions, resistivity often increases at low temperatures when $T$ is decreased. This $\sigma_{2D} \propto \ln T$ variation is due to weak localization, and was first experimentally demonstrated by Dolan and Osheroff [35].

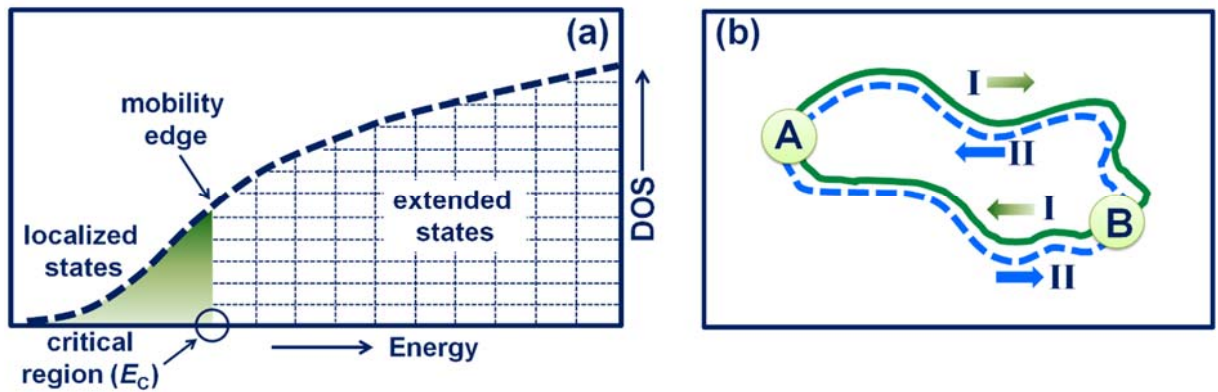

**FIGURE 4.** (a) Concept of the mobility edge $E_C$; electronic states > $E_C$ are extended and those states < $E_C$ are localized. If the Fermi energy is in the localized state, then the system is insulating; otherwise it is metallic. DOS: density of states. (b) Propagation of electronic waves in a disordered medium. If two waves follow the same path from A to B and then to A, one in a clockwise direction and the other in a time-reversed direction, then they interfere constructively on returning to A.

In 1960, Ioffe and Regel realized that the electron mean free path $\ell$ can never be shorter than the lattice spacing $a$, because at $\ell \leq a$, coherent quasi-particle motion would vanish and the system would be an insulator [36]. In 1972 Mott proposed a similar argument and formed a criterion $\ell_{min} \approx a$ for minimum metallic conductivity, [37]. This criterion was named the Mott-Ioffe-Regel (MIR) limit for resistivity saturation [38, 39]. Unfortunately, the MIR limit is not universally observed and disagrees with the now-widely-accepted scaling theory of localization. Nevertheless, the MIR limit is still quoted frequently and is now generalized to more complex media; the criteria ranging from $k_F\ell_{min} \approx 1$ through $\ell_{min} \approx a$ to $k_F\ell_{min} \approx 2\pi$ where the wave vector $k_F = 2\pi/\lambda$. In some materials MIT occurs at the MIR limit. For two-dimensional materials, $k_F\ell$ is related to the ratio of quantum resistance $h/e^2$ to sheet resistance $R_{sheet}$ as $k_F l = \frac{h/e^2}{R_{sheet}} \approx \frac{26 \text{ k}\Omega/\square}{R_{sheet}}$ [40].

## 3. Recent results on 5*d* perovskite iridates

5*d* perovskites, in particular iridates, show many new interesting phenomena; examples include novel insulating states, exotic magnetism, spin-liquid behaviors, Weyl semimetals, and topological insulators. In this regard, RP series $Sr_{n+1}Ir_nO_{3n+1}$ have been widely investigated, and are probably the most widely studied iridate system. In this section, we review some of the recent results on 5*d* perovskite iridates.

### 3.1 Perovskites

The term *perovskite* was named for the discovery of $CaTiO_3$ in honor of Count Lev Alexander Von Perovski, a Russian mineralogist. An ideal perovskite structure has an $ABX_3$ stoichiometry. Although most of the perovskite are oxides (X = Oxygen), other forms like fluorides, halides, sulphides also exist in literature. Perovskite $ABO_3$ and, more generally, perovskite families including its variations accept a large number of transition metal elements of various size and valence into the B-sites (e.g. $A^{1+}B^{5+}O_3^{2-}$, $A^{2+}B^{4+}O_3^{2-}$, and $A^{3+}B^{3+}O_3^{2-}$), so the variety of transition metal oxides is unlimited: example include cuprates, manganites, ruthenates, nickelates, titanates, and recently iridates [41, 42]. Along with this elemental diversity, mutually interacting quantum degrees of freedom such as lattice, spin, charge, and orbital lead to numerous emergent physical properties in perovskite families [1-14]. The perovskite families of current interest, iridates can be classified as follows:

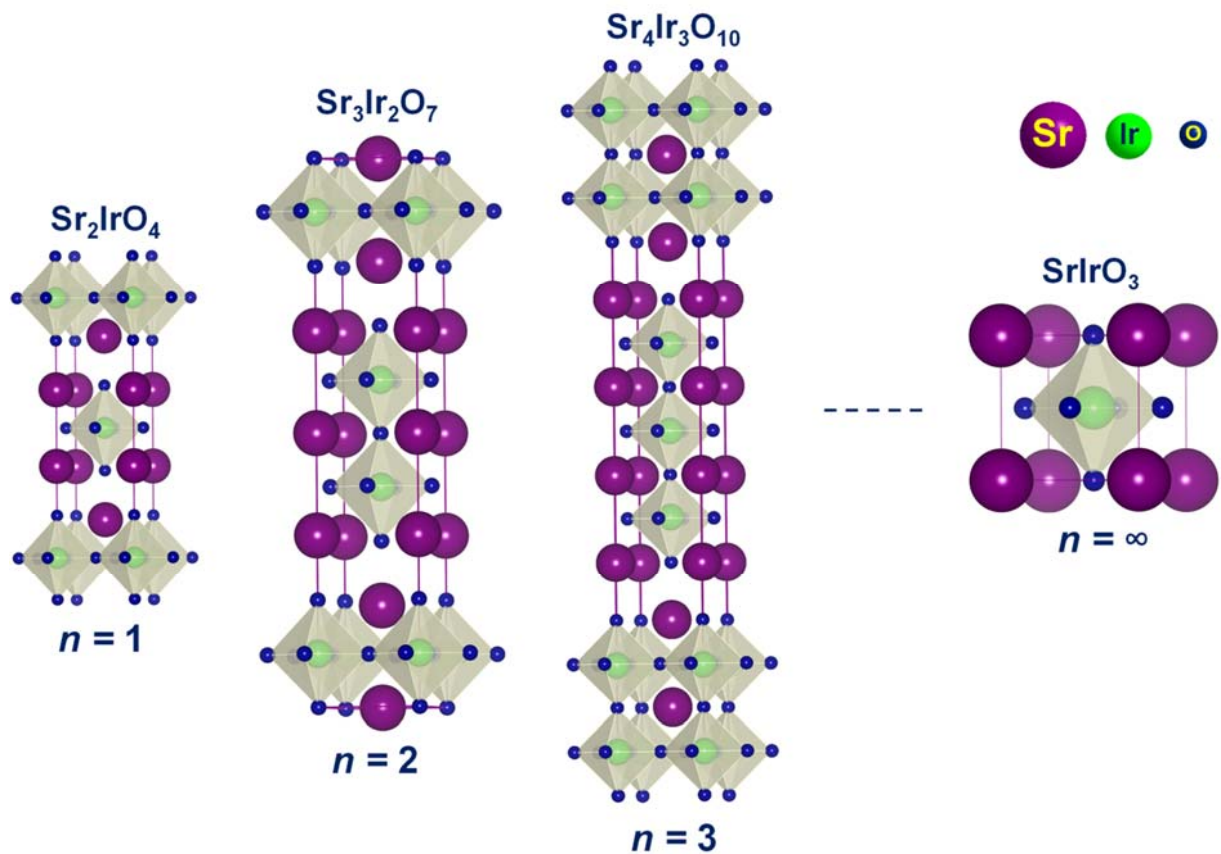

**FIGURE 5.** The $n = 1$ (Sr$_2$IrO$_4$), $n = 2$ (Sr$_3$Ir$_2$O$_7$), $n = 3$ (Sr$_4$Ir$_3$O$_{10}$), and $n = \infty$ (SrIrO$_3$) members of the homologous Ruddlesden–Popper series Sr$_{n+1}$Ir$_n$O$_{3n+1}$. Structures were drawn using VESTA software.

**(a) Perovskites**

General formula of ABO$_3$ where A or B are metal cations and O as anions. B cations are surrounded by six oxygen anions forming BO$_6$ octahedra. (e.g. SrIrO$_3$) (Figure 5).

**(b) Double Perovskites**

General formula of A$_2$BB'O$_6$ (ABO$_3$+AB'O$_3$) where the BO$_6$ and B'O$_6$ octahedra are alternatively arranged in two sublattices. For double perovskites, alkaline or rare earth ions occupy the A-sites, while the B-sites represent transition metal ions. (e. g. Sr$_2$RuIrO$_6$ constitutes alternative unit cells of SrRuO$_3$ and SrIrO$_3$)

**(c) Layered Perovkites:**

Layered perovskites consist of stacked two-dimensional slabs of the ABO$_3$ layer. Three subcategories have been recognized.

**(i) Ruddlesden-Popper Series**

The general formula of Ruddlesden-Popper (RP) series is $A_{n+1}B_nO_{3n+1}$, where $n$ represents the number of octahedral layers in the repeating unit, and can be visualized as repeated stacking of $AO(ABO_3)_n$ [43, 44]. $n = 1$ corresponds to one $BO_6$ octahedron ($Sr_2IrO_4$), and $n = 2$ to two $BO_6$ octahedra ($Sr_3Ir_2O_7$) (Figure 5). The differentiating characteristics for the layered perovskites are the motif ($Sr^{2+}$) that separates the layers, and the off-setting of the layers from each other. As $n$ increases from $n = 1$ to $n = \infty$, the dimensionality of the compounds changes from two to three.

**(ii) Aurivillius phase**

The general formula of the Aurivillius phase is $(Bi_2O_2)(A_{n+1}B_nO_{3n+1})$ [45]. This structure consists of layers of $Bi_2O_2$ separated by $n$ layers of perovskites

**(iii) Dion-Jacobson phases**

The general formula of the Dion-Jacobson phase is: $M(A_{n+1}B_nO_{3n+1})$ where M is a cation with valence +1, usually an alkali metal [46, 47].

Ideal perovskite $ABO_3$ has symmetric cubic structure with a lattice constant of □4 Å. However, most perovskites, deviate from the cubic structure if the Goldschmidt tolerance factor $t_f$, given by $t_f = \frac{r_A + r_O}{\sqrt{2}(r_B + r_O)}$, ($r_A$, $r_B$, and $r_O$ represents the ionic radii of ions A, B and O respectively) deviates from 1 [48]. When $t_f < 1$, (i.e., radius of cation A is small), the O anions move towards the A cation, so $BO_6$ octahedra tilt to shrink the available volume for A cations that is empty. Because B-O-B bonds are highly flexible and $BO_6$ octahedra are flexible in shape and size depending upon the cationic size, valence, and position, the overall deformation reduces the cubic symmetry, and result in structural transitions to orthorhombic, tetragonal, or hexagonal states that have lower symmetry than the cubic state.

**3.2 *d*-orbitals**

The critical factor in the orbital physics involving *d*-orbitals is their anisotropic charge distributions that arise from the wave functions, which take different shapes depending on energy when electrons are bound to atomic nuclei by Coulomb force. Electronic properties of perovskites would be severely affected by the chemistry of the transition metals at the center of $BO_6$ octahedra that have corner-sharing oxygen anions [15, 16]. Consider a transition metal element which is surrounded by six $O^{2-}$ in the $BO_6$ octahedron. This configuration gives rise to a crystal field which hinders the free motion of *d*-electrons; consequently, orbital angular momentum is usually quenched and the *d*-orbitals due to the crystal field energy $\Delta_{oct}$, split in energy into (1) $d_{x2-y2}$ and $d_{z2}$ states, which form two-fold degenerate $e_g$ orbitals with higher energy, and (2) three-fold degenerate $t_{2g}$ orbitals $d_{xy}$, $d_{yz}$, and $d_{zx}$ states at lower energy (Figure 6). As the degeneracy of spherical symmetry of an isolated atom is removed, the *d*-orbitals begin to fill, starting from a low energy state and continuing to higher energy states. The actual filling arrangement depends on the competition between the crystal field and on-site exchange interaction described by Hund's rule [49]. For example, for 5*d* perovskite $SrIrO_3$, $Ir^{4+}$ has five electrons in the *d*-orbitals ($5d^5$); the electrons distributions from basic viewpoint are illustrated (Figure 7). For transport properties, the band structure would depend sensitively on an overlap between the *d*-orbital of the B-site transition metal element and the *p*-orbitals of the surrounding oxygen's.

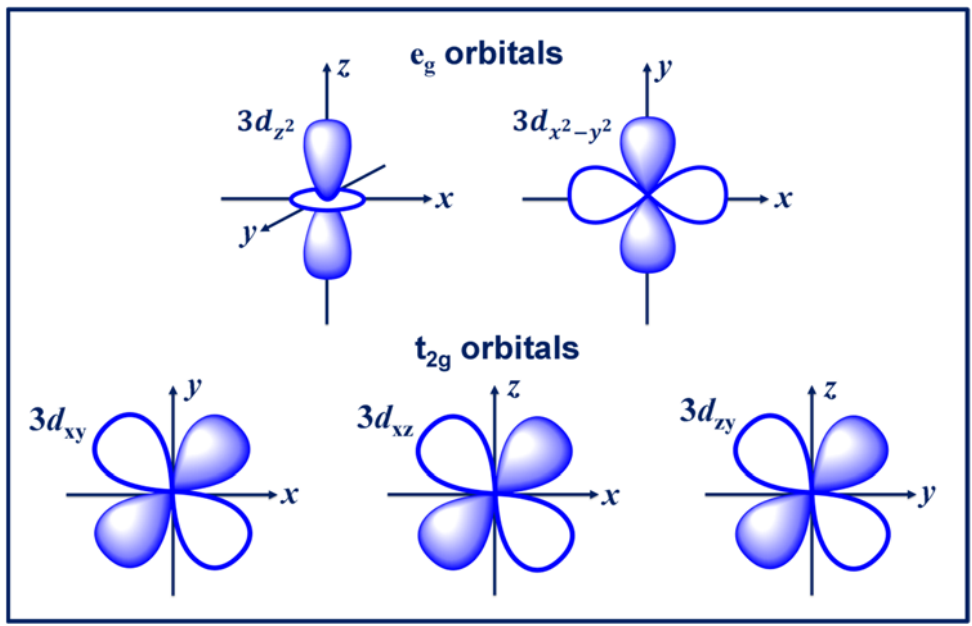

**FIGURE 6.** Five *d*-orbitals in a cubic crystal field which split into two $e_g$ orbitals and three $t_{2g}$ orbitals.

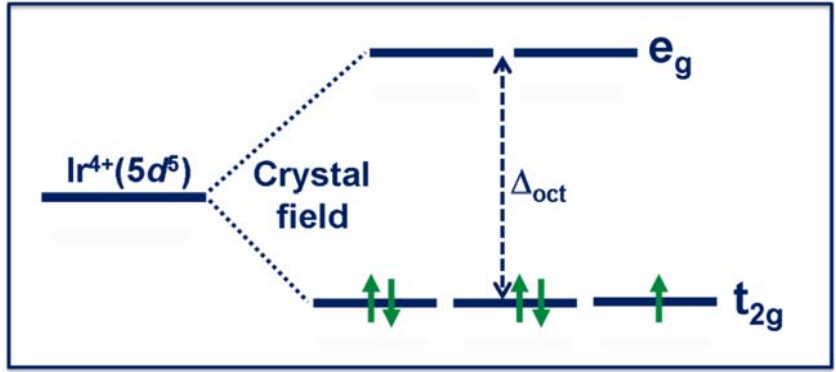

**FIGURE 7.** Simple view of distribution of five *d*-electrons of $Ir^{4+}$ ($d^5$) in $e_g$ and $t_{2g}$ orbitals splitted by octahedral crystal field energy $\Delta_{oct}$.

Generally, in transition metal oxides the electronic properties are further complicated by the interaction of various degrees of freedom surrounding the *d*-electrons, i.e., charge, orbital, spin, and lattice. These degrees of freedom would produce relevant energy scales of similar magnitude such as bandwidth, Coulomb repulsion, and SOC. Because of the large number of transition metals, the number of possible transition-metal-based perovskites is enormous. These transition metals include those of which the order orbital is 3*d* (e.g., Fe, Co, Ni), 4*d* (e.g., Mo, Ru, Rh), or 5*d* (e.g., W, Re, Ir). 3*d* orbitals are well-localized and thus form a narrow band (*W*) with a large on-site Coulomb interaction (*U*). 4*d* orbitals are spatially more extended than their 3*d* counterparts. As 5*d* orbitals are more spatially extended than 3*d* or 4*d* ones, as a result, nearest-neighbor orbitals overlap significantly, and therefore *W* is wider in 5*d* orbitals than in 3*d* or 4*d* cases, i.e., $W_{3d} < W_{4d} < W_{5d}$.

Because 5*d* orbitals are extended, on-site Coulomb repulsion or correlation *U* is weaker for 5*d* orbitals than form 3*d* or 4*d* ones, i.e., $U_{5d} < U_{4d} < U_{3d}$. Thus, naively one would expect 5*d* systems to be more metallic and less magnetic than those based on 3*d* and 4*d* transition metal oxides. Indeed, 5*d* perovskite $SrIrO_3$ is a correlated paramagnetic metal according to the expectation, but this is an exception. Surprisingly, many other 5*d* perovskites such as $Sr_2IrO_4$, and $Sr_3Ir_2O_7$ are insulators. This unexpected fact can be explained by the high SOC in 5*d* systems. In general, 5*d* systems have larger SOC than do 3*d* or 4*d* systems, i.e., $\Lambda_{3d} < \Lambda_{4d} < \Lambda_{5d}$ because SOC is proportional to the fourth power of the atomic number Z (i.e. $\Lambda_{soc} \propto$ $Z^4$) [49]. In iridate compounds, $Z = 77$ for Ir, so the SOC strength is very high, even comparable to on-site Coulomb repulsion (~0.5 eV). This high SOC strength leads to modification of the electronic structure and gives rise to novel emergent phenomena.

### 3.3 Ruddlesden–Popper series $Sr_{n+1}Ir_nO_{3n+1}$
### 3.3.1 Spin-orbit coupling and band structure evolution

In this section, we present some unusual phenomena that are observed in RP series $Sr_{n+1}Ir_nO_{3n+1}$ ($n$ = 1, 2, and ∞). Because the SOC strength is so much larger in Ir than in a typical 3$d$ system, SOC plays contributes to lifting the five-fold degeneracy of the atomic $d$-levels. In crystals, the crystal field and SOC act together to split the $t_{2g}$ levels into $J_{eff}$ = 3/2 and $J_{eff}$ = 1/2 levels (Figure 8). Two equivalent views of the splitting of 5$d$ orbitals due to $\Delta_{oct}$ and $\Lambda_{soc}$ are illustrated [17].

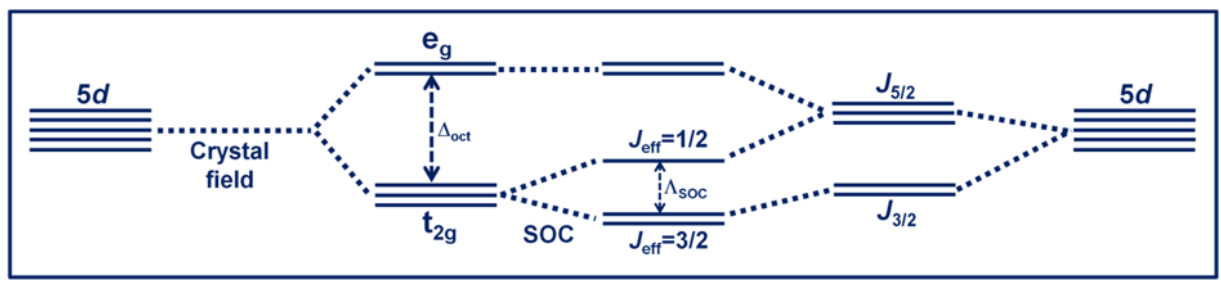

**FIGURE 8.** Two equivalent views of the splitting of 5$d$ orbitals due to crystal field $\Delta_{oct}$ and strong SOC $\Lambda_{soc}$. Horizontal lines in the center column represent the energy levels of the final configuration. The final configuration can be reached from the left side that depicts the crystalline view: crystal field splitting occurs first and then SOC generates $J_{eff}$ = 3/2 and 1/2 states. This result can also be viewed from the atomic viewpoint: the atomic five-fold degeneracy is lifted due to SOC, and then the crystal field turns quantum number $J$ into $J_{eff}$.

Now, considering electron hopping in solids, the $J_{eff}$ levels would become bands. $Ir^{4+}$ has five 5$d$ electrons; four of them fill the lower $J_{eff}$ = 3/2 band and one partially fills the $J_{eff}$ = 1/2 band with the Fermi level in the $J_{eff}$ = 1/2 band. The band structure evolves across the RP series $Sr_{n+1}Ir_nO_{3n+1}$ (Figure 9). In the $n$ = 1 case; $Sr_2IrO_4$; SOC splits the $t_{2g}$ band into two $J_{eff}$ bands, so the bandwidth of the conduction band ($J_{eff}$ = 1/2) is effectively reduced and correlation $U$ can split the conduction band again into upper Hubbard bands (UHB) and lower Hubbard bands (LHB). Thus, Sr$_2$IrO$_4$ becomes a Mott insulator (Figure 9(a)). As $n$ increases in the RP series, the bandwidth of the $J_{eff}$ = 1/2 band also increases, possibly due to increase in the coordination number. Sr$_3$Ir$_2$O$_7$, a barely insulator, is the intermediate case (Figure 9(b)). In SrIrO$_3$ with $n = \infty$, $U$ cannot split the conduction band with relatively wide bandwidth, so this compound is a correlated metal (Figure 9(c)).

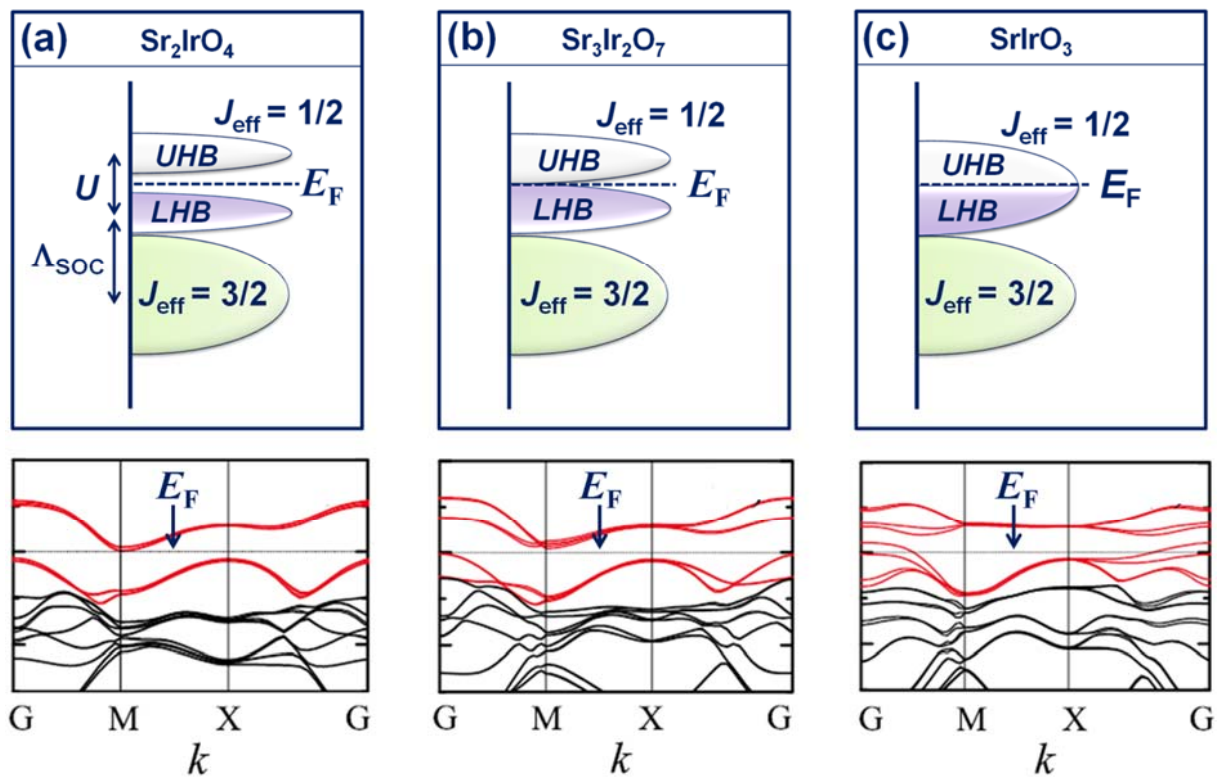

**FIGURE 9.** Evolution of the band structure across Ruddlesden–Popper series Sr$_{n+1}$Ir$_n$O$_{3n+1}$. Schematic diagrams (upper panel) and calculated results (lower panel, LDA+U+SOC). Sr$_2$IrO$_4$ ($n$ = 1) is a Mott insulator; SrIrO$_3$ ($n = \infty$) is a correlated metal. Barely insulating Sr$_3$Ir$_2$O$_7$ ($n$ = 2) is in intermediate. UHB: upper Hubbard; LHB: lower Hubbard band, respectively. In band structure calculation, the red and dark lines represent the $J_{eff}$ =1/2 and $J_{eff}$ =3/2 bands. The lower panels are reprinted with permission from [50]. © American Physical Society

### 3.3.2 Sr$_2$IrO$_4$

The first material in the RP series of $Sr_{n+1}Ir_nO_{3n+1}$ is a layered compound $Sr_2IrO_4$ ($n$ = 1). The crystal structure of $Sr_2IrO_4$ consists of alternating layered stacking of SrO-IrO$_2$-SrO perovskite units with ideal K$_2$NiF$_4$-type tetragonal cell (Figure 5). The lattice constants are ▫5.494 Å in the *ab*-plane and ~25.796 Å along the *c*-axis with space group *I4$_1$/acd* [51]. Most importantly, it has a highly-insulating nature (Figure 10(a)) [52] in contrast to a naive expectation that the extended nature of 5*d* orbitals would lead to a significant overlap of the nearest-neighbor orbitals and thus a broad electronic bandwidth. This insulating state arises from strong SOC and subsequent Coulomb repulsion (Figure 9(a)), and is termed a "Mott $J_{eff}$ = 1/2 insulating state" of the 5*d* electron system [53, 54]. Magnetically Sr$_2$IrO$_4$ exhibits weak ferromagnetism (canted antiferromagnetism) with $T_N$ ~ 240 K (Figure 10(b)) with very small ferromagnetic moment (0.023 $\mu_B$/Ir) [52]. The Mott insulating state due to strong SOC was first confirmed by Kim et al. by x-ray absorption (Figure 10(c)), based on the selection rules associated with the 2*p* to 5*d* transitions. Kim *et al.* also used theoretical calculations to identify the unusual nature of the Ir$^{4+}$ state in accord with [53]. Angle Resolved Photoemission Spectroscopy (ARPES) (Figure 10(d)) also confirmed the Mott $J_{eff}$ = 1/2 state [55].

Considering the origin of the Mott $J_{eff}$ = 1/2 insulating state, it should be very sensitive to external perturbations; if so the relevant energy parameters can be tuned to modify the electronic states. In Sr$_2$IrO$_4$ thin films under tensile (compressive) strain, the correlation energy is affected by in-plane lattice strain with increase (decrease) in bandwidth [56]. Sr$_2$IrO$_4$ remains an insulator even under pressure up to 55 GPa [57]; this stability illustrates the robustness of this insulating state. Also a pressure-induced, fully-reversible, giant piezoresistance was detected at room temperature [58]. The electronic band-gap could be tuned electrically, and this characteristic demonstrates potential application in next-generation electronic devices [59]. Alkali-metal doping induces a close resemblance of the electronic state to that of high-temperature cuprates, and therefore represents a step towards high-temperature superconductivity [60]. Many other exciting properties have been observed in this compound, including magnetic structural change, spin-orbit tuned MITs, lattice-driven magnetoresistivity, electron-doped tuned electronic structure, anisotropic magnetoresistance, and excitonic quasi-particle [61-66]. The interplay of crystal field splitting, SOC, and correlation effects in layered Sr$_2$IrO$_4$ determines the 5$d$ electronic structure and leads to realization of a completely new class of materials with a novel quantum state.

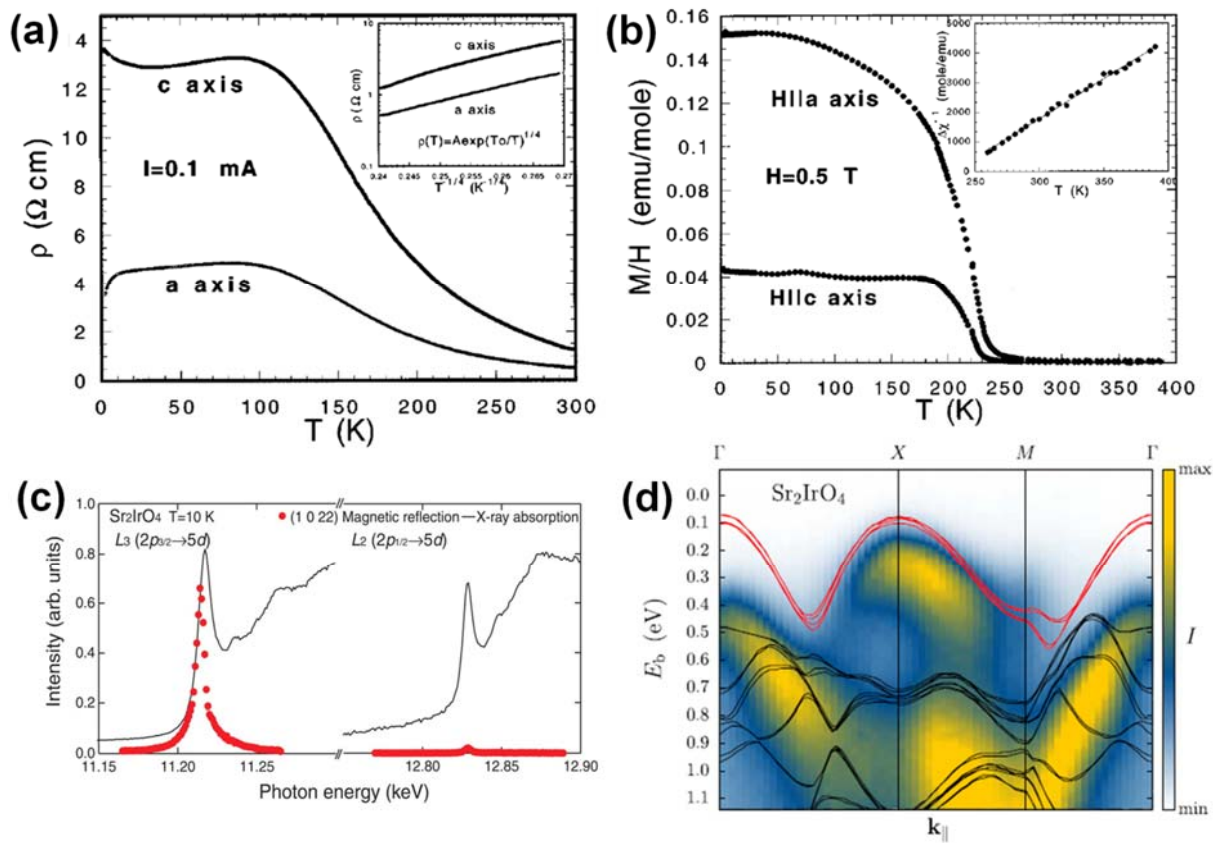

**FIGURE 10.** Characteristic properties of Sr$_2$IrO$_4$. (a) Resistivity along *ab*-plane and *c*-axis showing insulating nature with activation energy of ≈ 70 meV. (b) Magnetic measurements showing weak ferromagnetism with $T_N$ ~ 240 K. (c) Black lines: X-ray absorption spectra indicating the presence of Ir $L_3$ ($2p_{3/2}$) and $L_2$ ($2p_{1/2}$) edges around 11.22 and 12.83 keV. Red dots: intensity of the magnetic (1 0 22) peak. (d) Experimental ARPES spectra for Sr$_2$IrO$_4$ (hν = 85 *eV*; *T* = 100 K). $J_{eff}$ = 1/2 band (red line) and $J_{eff}$=3/2 band (black line) are shown. Reprinted with permission from [52, 54, 55]. © American Physical Society, © AAAS and © IOP Publishing.

### 3.3.3 Sr$_3$Ir$_2$O$_7$

The crystal structure of Sr$_3$Ir$_2$O$_7$ (*n* = 2) in this RP series of Sr$_{n+1}$Ir$_n$O$_{3n+1}$, is of tetragonal cell with *a* = 3.896 Å and *c* = 20.879 Å and space group *I4/mmm* [67]. It consists of strongly coupled bi-layers of Ir-O octahedra which are separated by Sr-O interlayer (Figure 5). With the increase of number of octahedral layers ($n$), the electronic bands progressively broaden and, in particular, the bandwidth of the $J_{eff}$ = 1/2 band increases from 0.48 eV for $n$ = 1 to 0.56 eV for $n$ = 2 [68]. Still the transport measurement shows a well-defined, barely-insulating $J_{eff}$ = 1/2 states (Figure 11(a)) [69]. Theoretical calculations based on LDA + $U$ + SOC also provide evidence for the existence of the barely-insulating band structure in which Fermi energy is between the $J_{eff}$ = 1/2 bands (Figure 9(b)) [50]. The onset of weak ferromagnetism occurs at $T_C$ ~ 285 K and is closely associated with the rotation of IrO$_6$ octahedra about the $c$-axis. Indeed, the temperature dependence M($T$) of magnetization closely tracks the rotation of the octahedra, as characterized by Ir-O-Ir bond angle. Sr$_3$Ir$_2$O$_7$ also exhibits an intriguing $M$ reversal for in-plane magnetization below 20 K with the onset of a rapid reduction at $T_D$ ~ 50 K (Figure 11(b)). The barely-insulating nature related to splitting of the $J_{eff}$ = 1/2 band was again observed with x-ray scattering and absorption (Figure 11(c)) [70], and supported by ARPES measurements (Figure 11(d)) [55].

The ground state of bi-layered Sr$_3$Ir$_2$O$_7$ is highly sensitive to small external perturbations such as chemical doping, high pressures, and magnetic field. By replacing Sr$^{2+}$ with La$^{3+}$, electrons can be doped into bulk samples, and can lead to an insulator-to-metal transition [71]. Also, application of a high hydrostatic pressure leads to a drastic reduction in the electrical resistivity; this observation suggests that the system is near an MIT [57]. Although this system has not been fully explored yet, some studies such as resonant inelastic x-ray scattering, scanning tunneling spectroscopy, optical conductivity, have been performed [72-74]. These observations indicated that Sr$_3$Ir$_2$O$_7$ is a good model system to explore the mechanism for novel Mott states near an MIT boundary at which competitive interplay between SOC and Coulomb interactions persists.

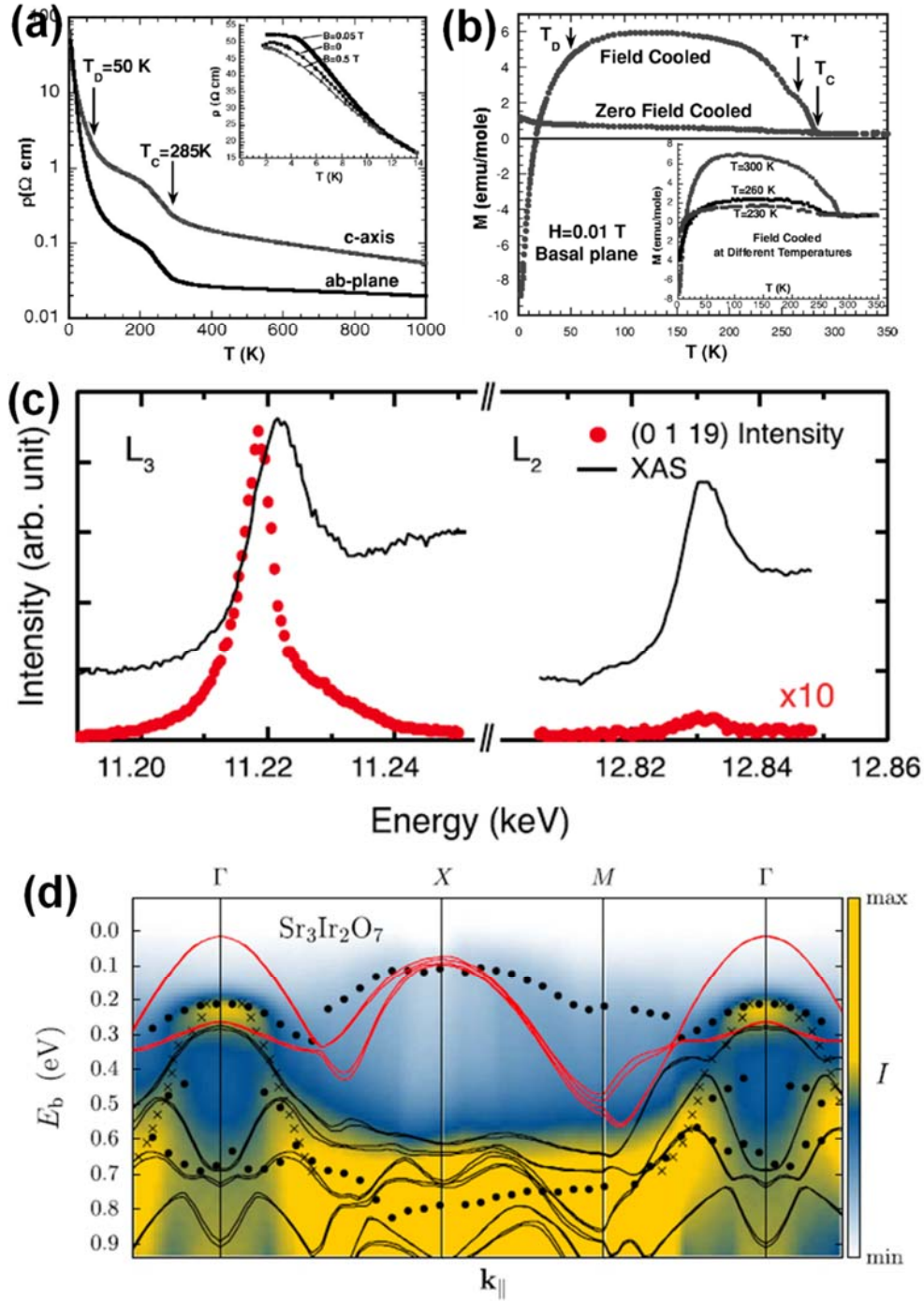

**FIGURE 11.** Characteristic properties of $Sr_3Ir_2O_7$. (a) Resistivity as a function of $T$ for the basal plane and along the $c$-axis. (b) In-plane magnetization $M$ vs. $T$ at magnetic field of 100 Oe. (c) Energy scan of (0 1 19) reflection scanned around Ir $L_3$ and $L_2$ resonances. Red dots: scattering intensity; black lines: x-ray absorption spectra. (d) Experimental ARPES spectra (hν = 10.5 $eV$; $T$ = 9 K). Calculated band structure: $J_{eff}$ = 1/2 (red line) and $J_{eff}$ = 3/2 bands (black line). Reprinted with permission from [55, 69, 70]. © American Physical Society, and © IOP Publishing.

### 3.3.4 SrIrO$_3$

The last compound in the RP series ($n = \infty$) is SrIrO$_3$; it is a correlated metal. SrIrO$_3$ is a three-dimensional system, and has the largest coordination number in the RP series. An increase in the coordination number would lead to the increase of the bandwidth, and thus correlation driven band splitting would not occur (Figure 9(c)). In fact, in SrIrO$_3$ the bandwidth of the $J_{eff}$ = 1/2 band reaches 1.01 eV, so no gap appears within the $J_{eff}$ = 1/2 band or between the $J_{eff}$ = 1/2 band and $J_{eff}$ = 3/2 band [50].

In 1971, Longo *et al.* first synthesized polycrystalline SrIrO$_3$ [75]. The stable ambient structure of SrIrO$_3$ is a monoclinic distortion of hexagonal BaTiO$_3$ structure [75, 76]. However, the structure transforms to an orthorhombic perovskite (space group *Pnma*) at 40 kbar and $T$ = 1000 °C (Table 2). If a perovskite sample obtained at high temperature and high pressure is quenched, it remains a perovskite at room temperature. Because this book is about perovskites, our main focus remains on the properties of perovskite SrIrO$_3$. Zhao *et al.*[77] and Blanchard *et al.*[78] again synthesized orthorhombic perovskite samples under high pressure and performed electric and magnetic measurements which showed that perovskite SrIrO$_3$ is truly a paramagnetic metal (Figure 12).

**TABLE 2.** Bulk SrIrO$_3$ can assume two forms depending on synthesis conditions. Corresponding lattice constants are summarized [75].

| Synthesis conditions | Structure | Lattice parameters |
|---|---|---|
| Atmospheric pressure, $T$ = 900 °C | Monoclinic distortion of hexagonal BaTiO$_3$ structure | $a$ = 5.604 Å  $b$ = 9.618 Å  $c$ = 14.17 Å  $β$ = 93.26° |
| $P_{O_2}$ = 40 kbar, $T$ = 1000 °C | Orthorhombic | $a$ = 5.60 Å  $b$ = 5.58 Å  $c$ = 7.89 Å |

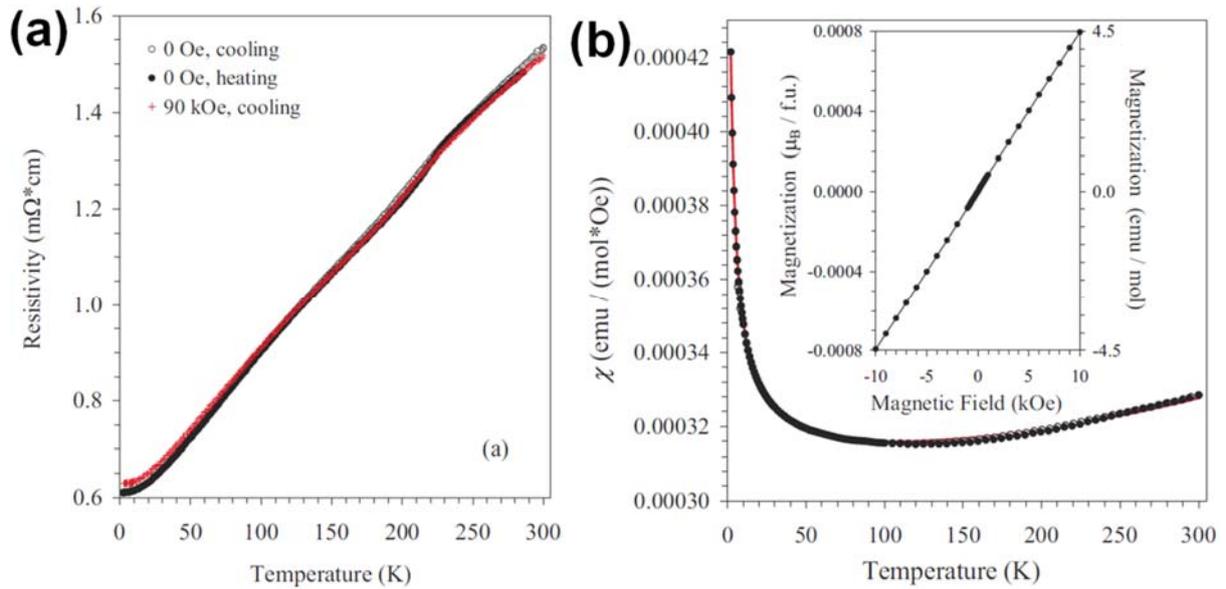

**FIGURE 12.** Characteristic properties of perovskite $SrIrO_3$. (a) Temperature dependence of electrical resistivity measured under different conditions. It remains metallic down to low temperatures. (b) It remains paramagnetic down to low temperatures. Magnetic susceptibility gives no sign of long range ordering. Inset confirms paramagnetism at $T$ = 5 K. Reprinted with permission from [78]. © American Physical Society.

In-situ ARPES showed that perovskite $SrIrO_3$ is an exotic narrow-band semi-metal [79]. The bandwidth was surprisingly narrower than other two-dimensional RP phases and the semi-metallic nature is caused by the unusual coexistence of heavy hole-like and light electron-like bands contrary to the coordination number argument given previously. The observed unusual property may originate from the interplay of strong SOC, dimensionality, and both in- and out-of-plane $IrO_6$ octahedral rotations. Recent theoretical calculations also suggest an extremely interesting possibility that the interplay of the lattice structure and large SOC produces Dirac nodes in the $J_{eff}$ = 1/2 band, and engineering topological phases at interfaces and in superlattices would alter the system to be close to a topological crystalline metal [80-82]. It is obvious that $SrIrO_3$ is an intriguing system in its own right and further studies are warranted.

## 4. Non-Fermi liquid physics and metal-insulator transitions in SrIrO$_3$ films

Among RP series compounds Sr$_{n+1}$Ir$_n$O$_{3n+1}$, the end member SrIrO$_3$ is of particular interest because it is a correlated metal that exhibits unusual electronic transport properties that deviate from the normal Fermi liquid behaviors. The MITs and associated non-Fermi liquid physics that occur in SrIrO$_3$ with strong SOC would provide an opportunity to extend the limit of our current knowledge of the physics of MITs as presented in the previous section. Thus, the present section constitutes the core of this chapter, and we first summarize salient features of the transport properties of SrIrO$_3$, particularly in thin film form. Then we proceed to present the relevant theoretical frameworks to understand the non-Fermi liquid physics that underlay the MITs in the system.

### 4.1 Salient features of the metal-insulator transitions in SrIrO$_3$

SrIrO$_3$ is believed to be close to an MIT as evidenced by the evolution of the RP series Sr$_{n+1}$Ir$_n$O$_{3n+1}$ from being an insulator to a correlated metal with increasing $n$ = 1 → 2 → ∞. The transport properties of correlated SrIrO$_3$ can be anticipated to be susceptible to external perturbations and MITs, and that the associated unusual properties could be induced in SrIrO$_3$ if, for example, an external stress is applied to the system. As introduced in the earlier section, the two most important mechanisms for MITs in correlated transition metal oxides are correlation-driven Mott localization and disorder-driven Anderson localization. When the system is under variable external stress, one may be able to tune *W, U, and D* to some extent and expose the interesting physics that is controlled by the parameters known as effective correlation (*U/W*) and effective disorder (*D/W*). For these reasons, we attempted to synthesize perovskite SrIrO$_3$. Perovskite SrIrO$_3$, however, is metastable at room temperature, and is obtainable only by applying an elevated pressure (~40 kbar) at high temperature (~1000 $^o$C) and subsequent quenching. While it is not easy to obtain single crystals of perovskite SrIrO$_3$ due to technical difficulties dealing with high pressures, the crystals can be stabilized by using thin film synthesis to produce them. In this case, the underlying substrates provide compressive strain, which replaces pressure, and epitaxial perovskite SrIrO$_3$ thin films are easily obtained. More importantly, compressive strains can be imposed on SrIrO$_3$ films by choosing substrates with appropriate lattice parameters. By depositing $SrIrO_3$ films on $GdScO_3$, $DyScO_3$, $SrTiO_3$, and $NdGaO_3$, one can impose progressively larger compressive strain in the films (Figure 13). Biswas *et al* [83] provides further details on thin-film synthesis.

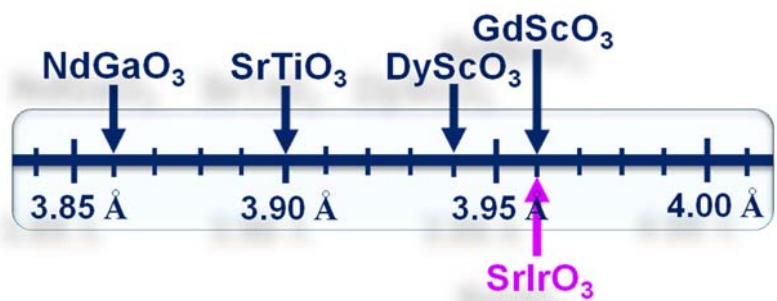

**FIGURE 13.** (Color Online) Pseudocubic lattice parameters of $SrIrO_3$ and various substrates available. $SrIrO_3$ and the substrates $GdScO_3$, $DyScO_3$, and $NdGaO_3$ are orthorhombic; $SrTiO_3$ is cubic.

The various scenarios on MITs suggest that an MIT may be obtained by varying the thickness of $SrIrO_3$ films. Indeed in films deposited on $GdScO_3$ which has a lattice constant well matched with that of $SrIrO_3$ (Figure 13), an MIT occurs as the thickness is reduced from 4 nm to 3 nm [83, 84]; a 4-nm film is metallic (with a resistivity upturn at low temperatures), but a 3-nm film is insulating. The resistivity upturn at low temperatures in the 4-nm film is well described by the weak localization in two dimensions, and the insulating behavior of the 3-nm film can be explained by the variable-range hopping. Thus, the thickness-driven MIT for $SrIrO_3$; (Figure 14(a)); is due to disorder, and falls in the class of Anderson localization [83]. Reducing the thickness of the film seems to increase the effectiveness of given disorder and of the grain size effect to scatter the charge carriers. The transition from being metallic with a low temperature upturn in resistivity for small disorder, to fully-insulating over the whole temperature range with the increase of disorder, is a realization of the disorder-driven MIT. However, the temperature variation of the resistance of all the metallic films (thickness 4 nm, 10 nm, and 35 nm) follows $\rho \propto T^{4/5}$. This nontrivial exponent indicates that although the MIT itself is driven by disorder, the underlying transport mechanism not so simple. Also in some materials, (e.g., $SrVO_3$), thickness-dependent MIT is caused by a decrease of the coordination number or by an increase in the correlation effect [85].

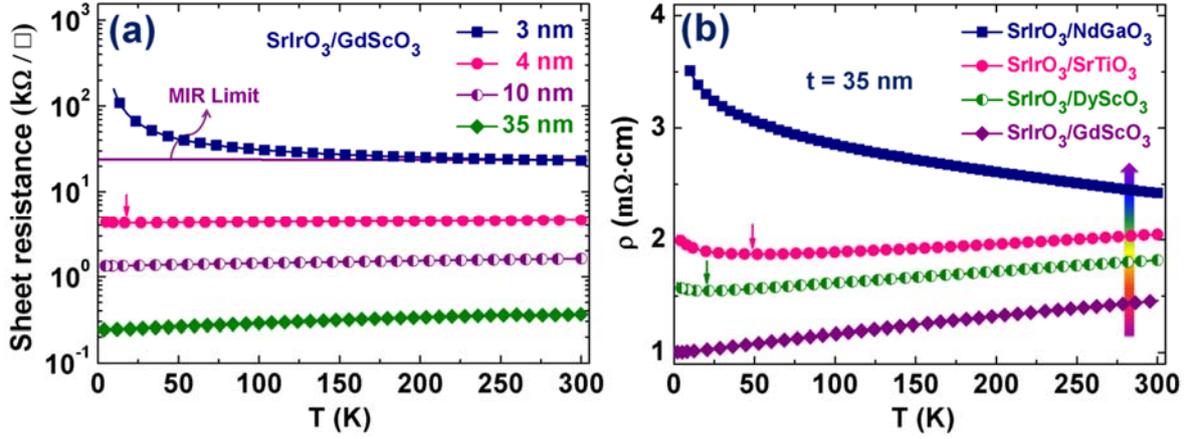

**FIGURE 14.** Metal-insulator transitions in perovskite SrIrO3. (a) Sheet resistance for films of varying thickness, and the thickness-driven metal-insulator transition. Mott-Ioffe-Regel (MIR) limit corresponds to the two-dimensional quantum sheet resistance $(h/e^2)$ ~26 kΩ/□. (b) Resistivity for films of constant thickness 35-nm on various substrates, and compressive-strain-driven metal-insulator transitions. The vertical arrow indicates an increase of compressive strain. Small arrows in both figures indicate the temperature below which resistivity upturns show up. Reprinted with permission from [83]. © 2014, AIP Publishing LLC.

MIT also occurs in perovskite SrIrO3 thin films when compressive strain is imposed on the films when the thickness is kept constant (Figure 14(b)) [83]. The imposed strain changes the Ir-O bond length $d$ and the Ir-O-Ir bond angle $\theta$; these changes affect the bandwidth as $W \propto \frac{\cos\psi}{d^{3.5}}$, where $\psi = (\pi - \theta)/2$ is the buckling deviation of the *Ir-O-Ir* bond angle $\theta$ from $\pi$ [83]. $d$ is relatively difficult to change, but $\theta$ can be readily affected by lattice strain. When $\theta$ is decreased by compressive strain, the electronic hopping integral between Ir 5*d* orbitals is reduced, so the buckling drives the system towards the insulating state. In addition, compression would also slightly increase correlation (*U*). Thus, the overall change in effective correlation (*U/W*) would seem to induce the MIT, however, as compressive strain increases, the effective disorder (*D/W*) in the system also increase. Disorder was already shown to play a decisive role in the thickness-driven MIT, and has a strong influence in the strain-driven MIT, as exposed by optical conductivity measurements [86]. Optical absorption spectroscopy provides insight into electronic band structure and free-carrier dynamics; optical absorption spectroscopy measurements of compressively strained $SrIrO_3$ films showed Drude-like, metallic responses without an optical gap opening. This result indicates that localization has a measurable effect on strain-induced MIT in perovskite $SrIrO_3$ thin films. In fact, extensive transport measurements in the compressively-strained $SrIrO_3$ revealed is that the MIT is not simply due to either disorder or correlation [83, 87]. Thus, $SrIrO_3$ seems to provide a rare example in which the interplay of correlation and disorder in the presence of SOC causes the MIT.

The most remarkable feature of the electrical transport of perovskite $SrIrO_3$ under compression is that the temperature variation of resistivity deviates from the Fermi liquid behavior in a peculiar way. For $SrIrO_3$ films of 35-nm thickness on $GdScO_3$, $DyScO_3$, $SrTiO_3$, and $NdGaO_3$ substrates, the electrical resistivity not only shows non-Fermi liquid behaviors ($\rho \propto T^\varepsilon$ with $\varepsilon \neq 2$) but $\varepsilon$ evolves from 4/5 to 1 to 3/2 as the compressive strain is increased; specifically, $\rho \propto T^{4/5}$ for films on $GdScO_3$, $\rho \propto T$ for films on $DyScO_3$, and $\rho \propto T^{3/2}$ for films on $SrTiO_3$ (Figure 15) [83]. Films on $NdGaO_3$ are subjected to the largest strain, and become insulating (Figure 14(b)). The present strain-driven MIT is clearly contrasted to the thickness-driven MIT, during which $\varepsilon$ remains constant at 4/5 as the thickness is reduced and the system approaches the MIT [83].

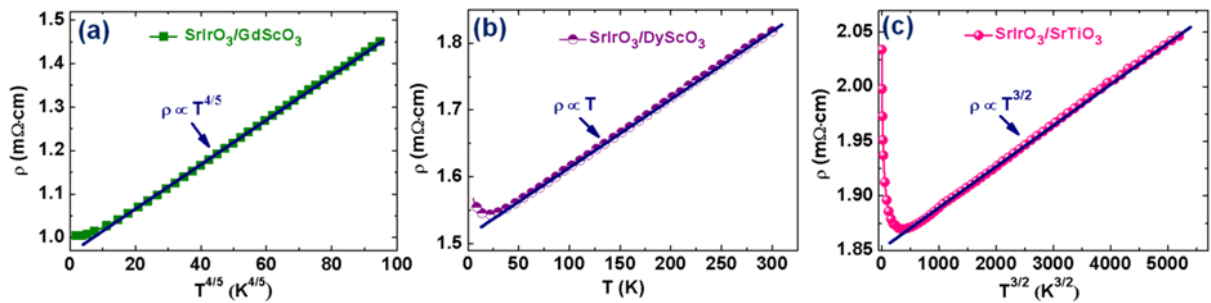

**FIGURE 15.** Temperature-dependent resistivity of $SrIrO_3$ thin films of thickness 35 nm on various substrates. The films show distinctly non-Fermi liquid behaviors as $\rho \propto T^\varepsilon$ with: (a) $\varepsilon$ = 4/5 for films on $GdScO_3$, (b) $\varepsilon$ = 1 for films on $DyScO_3$, and (c) $\varepsilon$ =

3/2 for films on SrTiO$_3$. The films on DyScO$_3$ and SrTiO$_3$ show resistivity upturns at low temperatures. Adapted with permission from [83]. © 2014, AIP Publishing LLC.

The cause of these peculiar non-Fermi liquid behaviors and consequent MIT in the strain-driven case is not clear. Arguably, because SrIrO$_3$ is paramagnetic without long-range magnetic ordering, localized states might induce formation of local magnetic moments without collective magnetic fluctuations because they occur near the MIT. Such localized moments or small magnetic clusters can influence the electronic transport significantly in the presence of disorder. Indeed, the presence of disorder in the strained films is indicated by the increases in resistivity at low temperatures (Figure 14). Also disorder and the evolution of non-Fermi liquid physics are possibly inter-connected. In a correlated metal, when disorder is sufficiently high, the system can enter a so-called Griffiths phase, which consists of a mixture of islands of Fermi liquid and Mott insulating regions, and that has non-Fermi liquid behaviors. In the following sections we will try to propose a model that incorporates all these factors, and that may explore a new paradigm for non-Fermi liquid physics.

**4.2 UV physics: Emergence of localized magnetic moments and dynamical mean-field theory**

Essential experimental features for SrIrO$_3$ films of thickness 35-nm grown on GdScO$_3$, DyScO$_3$, SrTiO$_3$, and NdGaO$_3$ are non-Fermi liquid transport phenomena near MIT, where electrical resistivity $\rho \propto T^{\varepsilon}$ show anomalous temperature dependences with $\varepsilon$ = 4/5, 1, 3/2, and -1/4 respectively. (The resistivity for the insulating phase can be described with a negative exponent.) The continuous change of the temperature exponent $\varepsilon$ implies that a particular type of interplay between correlations of electrons and disorders are expected have an important influence on physics near the MIT, and raises fundamental questions about the nature of the non-Fermi liquid. The first question is whether this non-Fermi liquid physics is involved with either UV (ultraviolet) or IR (infrared) physics. Here, UV physics means that localized magnetic moments appear near an MIT, and are regarded to be the source of strong inelastic scattering events due to their extensive entropy and to be responsible for non-Fermi liquid transport phenomena. The theoretical framework of dynamical mean-field theory was designed to simulate this local-moment physics quite well [88].

Dynamical mean-field theory describes Mott quantum criticality successfully, and suggests that it appears at high temperatures, where not only MITs but also the so-called bad metal physics have been revealed [89, 90]. A noticeable point is that this local-moment UV physics seems to be universal, regardless of IR physics in which such local moments are expected to disappear by, forming either singlets or magnetic orders and thereby reducing the entropy dramatically at low temperatures. Because incompletely-screened local moments have important effects in non-Fermi liquid physics, the appearance of negative magnetoresistance (MR) (MR=$\frac{\rho(B)-\rho(0)}{\rho(0)}$) can be expected. However, the experiment on 35-nm $SrIrO_3$ films on various substrates confirms positive MR, which is less than 1% up to magnetic field strength of 9 Tesla (Figure 16) [83]. Although this result does not necessarily mean that the local-moment physics may not be important in the MIT of $SrIrO_3$ thin films, the positive MR leads us to focus on IR physics.

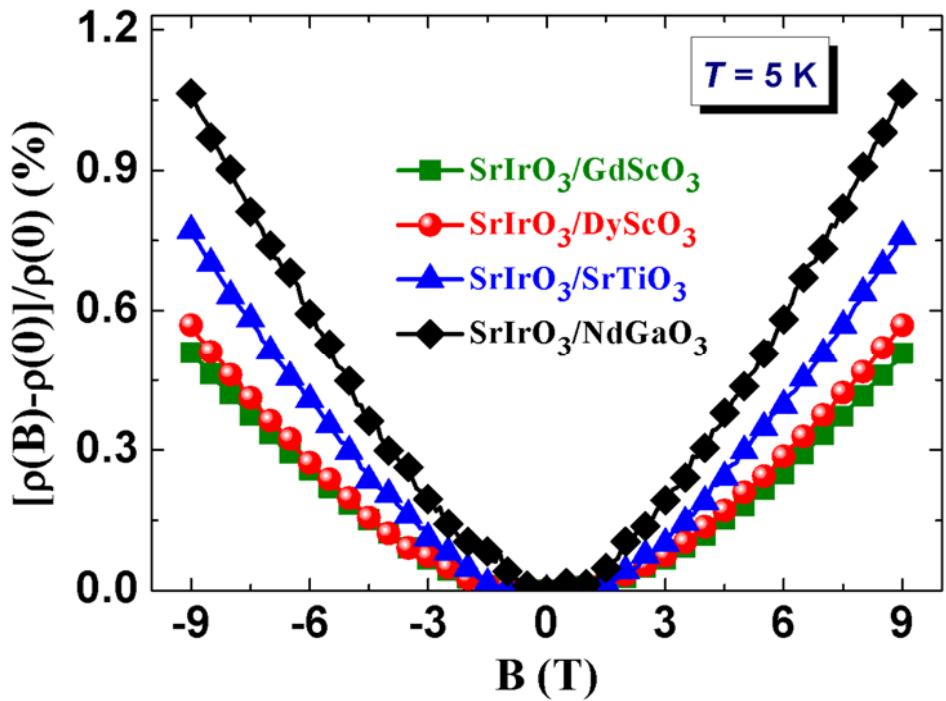

**Figure 16.** Magnetoresistance (MR) at $T$ = 5 K of $SrIrO_3$ thin films of thickness 35 nm grown on various substrates. MR was positive in all films regardless of compressive

strain which determines the magnitude of MR

Another important phenomenon is that spectral weight transfer in optical conductivity from the Drude part to a mid-infrared region as temperature increases [38]. This change is a characteristic feature of bad metals, and occurs at finite temperatures in the metallic side near an MIT. Dynamical mean-field theory reveals that local-moment UV physics describes this bad-metal Mott physics nicely. If local-moment physics are not considered, the spectral weight transfer in the optical conductivity cannot be easily reproduced without symmetry-breaking. Unfortunately, no experiment has been performed yet in determining the importance of UV local-moment physics on non-Fermi liquid states of SrIrO$_3$ thin films. With the contradictory picture of UV physics and observed positive MR, in the next section we focus on IR physics, which assumed to be responsible for the observed non-Fermi liquid physics near the MIT of SrIrO$_3$ thin films.

**4.3 IR physics: Slater quantum criticality vs. Mott-Anderson-Griffiths scenario**

Here, IR physics means that long-wave length and low-energy fluctuations determine the non-Fermi liquid physics near the MIT of SrIrO$_3$ thin films. Then, quantum criticality would be the first choice, where quantum critical fluctuations involved with symmetry breaking which cause to strong inelastic scattering between low-energy electrons. This scattering is responsible for non-Fermi liquid transport phenomena [91]. Unfortunately, we failed to figure out quantum criticality involved with any kinds of orders, in particular, those associated with magnetism, although we do not exclude more-delicate symmetry breaking near the MIT of SrIrO$_3$ thin films. Furthermore, quantum criticality itself cannot readily explain the continuous change of $\varepsilon$ in the relationship $\rho \propto T^{\varepsilon}$ of electrical resistivity to temperature on the metallic side of the MIT. One way to understand this continuous change is to consider the effect of disorder on quantum criticality. The Harris criterion is on the stability of quantum criticality against weak randomness, in the sense of average, involved with a space dimension and a critical exponent of correlation length [92]. When the Harris criterion is violated, the clean quantum critical point becomes destabilized. As a result, a novel disordered quantum critical point can emerge, respecting the Harris criterion. However, the continuous evolution of the non-Fermi liquid physics is difficult to understand even in this situation. However, the strength of randomness can grow indefinitely, resulting in the so-called infinite randomness fixed point; when this happens samples become extremely inhomogeneous due to effectively enhanced disorders, with ordered regions coexisting with disordered islands. The statistical distribution of the ordered islands shows a power-law tail, which implies that rare events that correspond to the power-law tail of the distribution function have important influences in non-Fermi liquid physics, particularly, thermodynamics [93]. For example, an antiferromagnetic Heisenberg model with random exchange coupling lies at an infinite randomness fixed point, at which the antiferromagnetic quantum critical point disappears and is replaced by a random singlet state [94]. An attractive feature of the infinite randomness fixed point is that non-Fermi liquid physics still survive near it, i.e., away from the quantum critical point, where dynamics of rare regions associated with the long tail part of the distribution function governs singular behaviors of the extremely inhomogeneous state, These phenomena are referred to as quantum Griffith effects [95]. The continuous change of non-Fermi liquid transport exponents reminds us of the quantum Griffith phase.

We suspect that the MIT in 35-nm $SrIrO_3$ films on $GdScO_3$, $DyScO_3$, $SrTiO_3$, and $NdGaO_3$ can be achieved by electron correlations, in which lattice mismatches between substrates and $SrIrO_3$ thin films are expected to control the ratio between interactions and hopping integrals. An essential question is on the role of disorder in this Mott transition, combined with electron correlations. Considering that the films show positive MR at magnetic field strength up to 9 Tesla, we suggest that magnetic correlations may not have important influence in this MIT. In fact, the positive MR suggests that the resulting insulating phase may be paramagnetic.

Furthermore, the residual resistivity is close to the Mott-Ioffe-Regel (MIR) limit; this similarity implies that the concentration of disorders is not low. Therefore, we conjecture that the interplay between electron correlations and disorders can give rise to a Griffith-type phase between Landau's Fermi liquid state and the Mott-Anderson insulating phase; the Griffith-type phase allows the continuous change of transport exponents. We call this physics the Mott-Anderson-Griffith scenario. Until now, the Griffith scenario has been realized near the infinite randomness fixed point [93], at which extreme inhomogeneity and associated rare events are responsible for non-Fermi liquid physics that have varying critical exponents. Although the mechanism by which such an infinite randomness fixed point can appear in the Mott-Anderson transition has not been identified, fluctuations between metallic and insulating islands as rare events are expected to allow development of the Mott-Anderson-Griffith phase.

### 4.4 Model Hamiltonian and tentative Global phase diagram

Recalling the interplay between the spin-orbit coupling and the Hubbard interaction, we start from an Anderson-Hubbard model with one band (Eq. (2));

$$H = -t \sum_{ij\sigma} c_{i\sigma}^\dagger c_{j\sigma} + U \sum_i \left( \sum_\sigma c_{i\sigma}^\dagger c_{i\sigma} \right)^2 - \sum_i v_i \left( \sum_\sigma c_{i\sigma}^\dagger c_{i\sigma} \right) \quad (2)$$

where, $c_{i,\sigma}$ represents an electron field at site $i$, and $\sigma$ expresses a Kramers doublet state given by total angular momentum, $t$ is a hopping integral, $U$ is the strength of on-site Coulomb interaction, and $v_i$ is a random potential introduced by disorder. When electron correlation becomes negligible, the Hamiltonian reduces to the Anderson model, showing a continuous phase transition from a diffusive Fermi-liquid state to an Anderson insulating phase in three dimensions. Local density of states (LDOS) occur (Figure 17), in which localized eigenstates are given by discrete energy levels. This condition occurs below the mobility edge in the diffusive Fermi-liquid state, and over the whole range of energy in the insulating phase. When contributions of random impurities can be neglected, the Hamiltonian becomes the Hubbard model, which shows a Mott transition from Landau's Fermi-liquid state to a paramagnetic Mott insulating phase. Regarding this Mott transition based on the UV local-moment physics, yields first order MIT [88]. In contrast, using IR physics to explain the Mott transition yields a continuous MIT, which will be discussed below. In the Landau's Fermi liquid state a coherent peak occurs at the Fermi energy in the electron spectral function besides incoherent humps and it disappears to transfer into an incoherent background, resulting in upper and lower Hubbard bands in the paramagnetic Mott insulating phase. A further question is whether the Griffith-type phase appears near the Mott-Anderson transition in the middle region of the phase diagram (Figure 17)

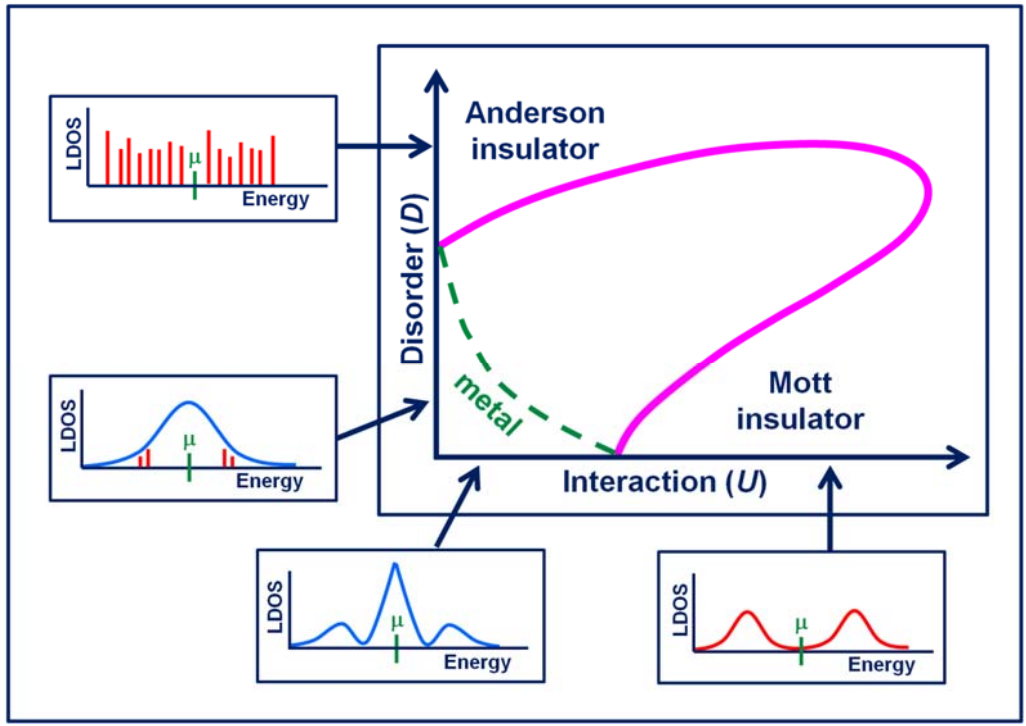

**FIGURE 17.** A schematic phase diagram of Anderson-Hubbard model with one band and local density of states (LDOS) of each phase. Adapted with permission from [96]. © 2005 American Physical Society.

As discussed, we try to describe us IR physics to explain the Mott-Anderson MIT in SrIrO$_3$ thin films. Incompletely-screened local moments can be expected to be screened completely at low temperatures. One mechanism for this screening is the Kondo effect, in which localized magnetic moments form singlets with itinerant electrons, well described by dynamical mean-field theory. The other mechanism entails localized magnetic moments form singlets with themselves as a RKKY-type (Ruderman, Kittel, Kasuya, Yosida) interactions [97-99]. The resulting paramagnetic Mott insulating phase is called a spin-liquid state, in which charge-neutral spinons may emerge as low-energy elementary excitations, and may interact among themselves through abundant singlet fluctuations, referred to as gauge fields [100]. Recalling the positive MR, we discuss the Mott-Anderson transition based on the spin-liquid physics.

## 4.5 Generalization of Finkelstein's nonlinear sigma model approach near the Mott-Anderson transition

To describe the Mott transition involved with the spin-liquid physics, we take the U(1) slave-rotor representation [101] as $C_{i\sigma} = e^{-i\theta_i} f_{i\sigma}$, in which an electron field is expressed as a composite operator of charge and spin degrees of freedom. $\theta_i$ accounts for the dynamics of collective charge fluctuations (sound modes) and $f_{i\sigma}$ expresses a charge-neutral spinon field for collective dynamics of spin degrees of freedom. Then, an effective mean-field theory for such variables of $\theta_i$ and $f_{i\sigma}$ can be easily formulated from the Hubbard model without randomness; in this model the Hubbard Hamiltonian is decomposed into two sectors describing the dynamics of spinons and zero-sound modes, respectively, given by (Eq. (3) and Eq. (4) )

$$S_F = \int_0^\beta d\tau \left[ \sum_{i,\sigma} f_{i\sigma}^\dagger (\partial_\tau - \mu) f_{i\sigma} - t\chi_f \sum_{\langle ij\rangle,\sigma} (f_{i\sigma}^\dagger f_{j\sigma} + h.c.) \right] \quad (3)$$

$$S_B = \int_0^\beta d\tau \left[ \frac{1}{2U} \sum_i (\partial_\tau b_i^\dagger)(\partial_\tau b_i) - t\chi_\theta \sum_{\langle ij\rangle} (b_{i\sigma}^+ b_{j\sigma} + H.c.) + \lambda \sum_i (|b_i|^2 - 1) + 2L^2 zt\chi_f \chi_\theta \right] \quad (4)$$

Here, the conventional saddle-point approximation has been performed for a spin-liquid-type Mott insulating phase. $\chi_f$ describes band renormalization for electrons and $\chi_\theta$ approximately expresses the width of incoherent bands. $\lambda$ is a Lagrange multiplier field to control the spin-liquid to Fermi-liquid phase transition, and is regarded to be the chemical potential of bosons. This results from a nonlinear $\sigma$−model description, in which the rotor variable $e^{-i\theta_i}$ is replaced with $b_i$ and the unimodular constraint $|b_i|^2 = 1$ is considered. $z$ is the nearest coordinate number of our lattice, and $L^2$ is the size of the system.

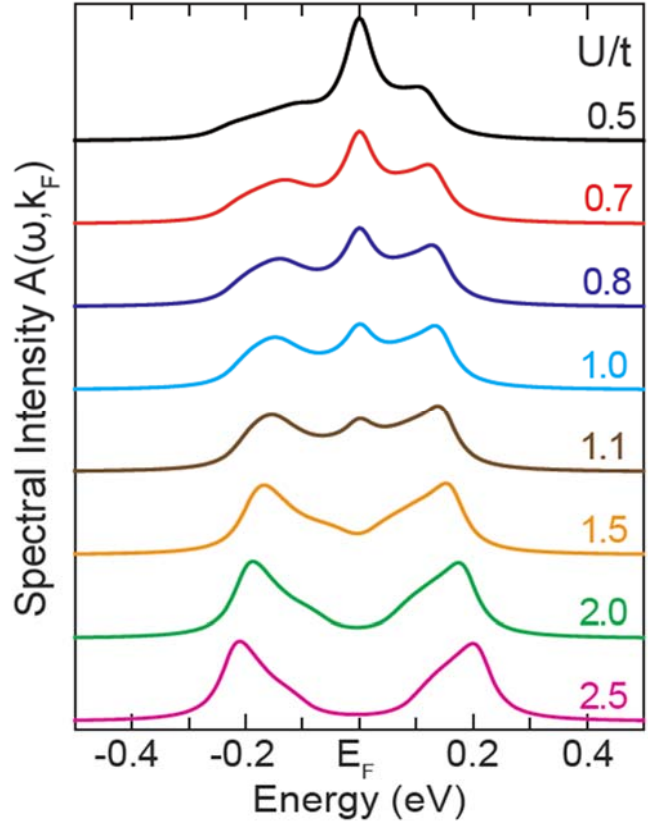

**FIGURE 18.** Evolution of the electron spectral function from Landau's Fermi-liquid state to a spin-liquid Mott insulating phase [102].

The Mott transition in this spin-liquid approach is realized by the condensation transition of $b_i$, by which zero-sound modes are gapless in the Fermi liquid state, i.e., $\langle b_i \rangle \neq 0$ but become gapped in the spin-liquid phase, i.e., $\langle b_i \rangle = 0$. As a result, the height of the coherent peak in the electron spectral function, proportional to the condensation amplitude $|\langle b_i \rangle|^2$, decreases gradually to disappear toward the spin-liquid Mott insulating phase; during this process the spectral weight is transferred to the incoherent background of upper and lower Hubbard bands (Figure 18). Our problem is to introduce a random potential into this spin-liquid Mott transition. As long as the strength of the random potential remains smaller than the spinon bandwidth, we are allowed to deal with the role of the random potential perturbatively. This renormalization group analysis has been performed to reveal that the clean spin-liquid Mott critical point becomes unstable as soon as the random potential is turned on. As a result, a disorder critical point appears to be identified with a transition from diffusive spin-liquid glass insulator to diffusive Fermi-liquid metal, in which the diffusive spin-liquid glass state consists of a diffusive spin-liquid phase of spinons and a charge glass phase of sound modes (Figure 19) [103].

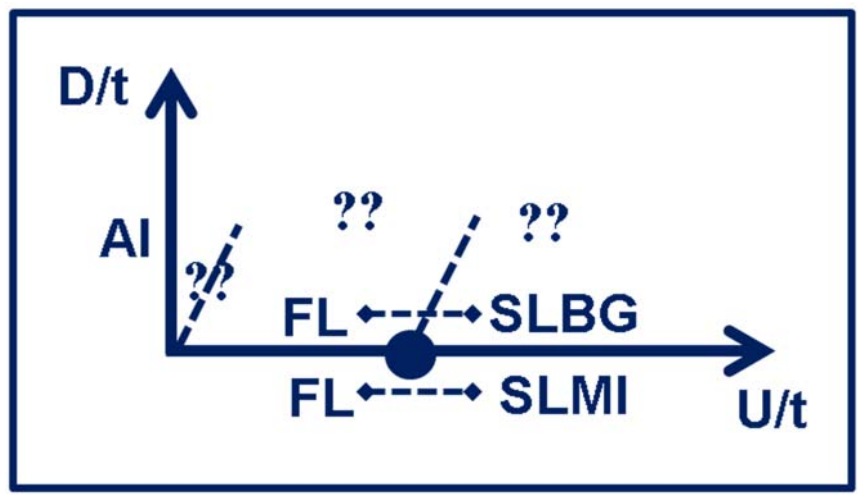

**FIGURE 19.** A schematic phase diagram for the spin-liquid Mott-Anderson transition based on weak-coupling renormalization group analysis. FL: Fermi liquid metal, SLMI: spin liquid Mott insulator, SLBG: spin liquid Bose glass, AI: Anderson insulator. Bose glass means that charge dynamics described by sound modes displays glassy behaviors. *U*: strength of local interactions, *D*: disorder and *t*: hopping integral of electrons. Adapted with permission from [103]. © 2006 American Physical Society.

In this renormalization group analysis, the spinon conductivity has been used as an input parameter, rather than being self-consistently determined, and the diffusive dynamics of spinons has been assumed. As a result, the renormalization group study reached the conclusion that the weak-disorder quantum critical point depends on the residual spinon conductivity, which means that the universality may not appear around this disorder quantum critical point. Suppose that $SrIrO_3$ thin films on the substrate of $GdScO_3$ are near this disordered MIT. Since $SrIrO_3$ thin films on $DyScO_3$, $SrTiO_3$, and $NdGaO_3$ substrates are also near this quantum critical point and their spinon conductivities differ from each other, we may observe continuous change of the temperature exponent $\varepsilon$ of the electrical resistivity. Unfortunately, however, this previous study does not evaluate the spinon conductivity self-consistently, so the discussion cannot be beyond our speculation. Therefore a theoretical framework must be developed to determine both transport coefficients of spinons and sound modes self-consistently.

One problem is that the previous renormalization group analysis does not consider effective interactions between diffusions and Cooperons, i.e., weak-localization corrections [34]. The replica nonlinear $\sigma-$model approach serves a natural theoretical framework to introduce such quantum corrections [104, 105]. By introducing the replica trick into the effective action for spinons and sound modes, and taking the diffusive spin-liquid fixed point as a saddle point for the spinon dynamics, one can derive an effective field theory as follows (Eq. (5)) [106]:

$$\begin{aligned} S_{eff} = \int_0^\beta d\tau \int d^d r \ \Bigg\{ &\frac{\pi N_F^f}{4}\sqrt{-\partial_\tau^2} tr(KQ) + \frac{\pi N_F^f}{4} D_c tr(\nabla Q - ia^c[Q,\tau^3])^2 \\ &+ \frac{1}{U}\{(\partial_\tau + i\varphi)b^\dagger\}\left(1 - U\frac{\sqrt{-\partial_\tau^2}}{\sqrt{-\partial_\tau^2 + D_c(-\nabla^2)}}\right)\{(\partial_\tau - i\varphi)b\} \\ &+ t\eta_\theta|(\nabla - ia^c)b|^2 + \lambda_c b^\dagger b \Bigg\} \\ &+ \frac{1}{\beta}\sum_{i\nu}\sum_q a_i^c(q,i\nu)(\gamma_a|\nu| \\ &+ v_a^2|q|^2)\ P_{ij}^T a_j^c(-q,-i\nu) \end{aligned} \tag{5}$$

Dynamics of diffusions and Cooperons of the spinon sector are described by the replica nonlinear $\sigma-$model of the first line. $Q$ is a $4N \times 4N$ matrix field, located on the Grassmannian manifold of $\frac{Sp(4N)}{Sp(2N)\times Sp(2N)}$, which spans retarded and advanced, Kramers doublet, and $N$ replica spaces. $K$ is a constant diagonal $4N \times 4N$ matrix, given by $I_{2N\times 2N}$ for the retarded part and $-I_{2N\times 2N}$ for the advanced sector. $N_F^f$ is the spinon density of states at the spinon Fermi-energy and $D_c$ is the diffusion coefficient. $a^c$ a gauge field to describe singlet excitations, more precisely, spin-chirality fluctuations that are neglected in the previous mean-field analysis but introduced here beyond the saddle-point approximation [100], in which the dynamics of gauge fluctuations is described by the last term, which is given by the polarization function of the diffusive dynamics of spinons. When the effects of gauge fluctuations in the dynamics of diffusions and Cooperons can be neglected, the replica nonlinear $\sigma$ −model reduces to that for the Anderson localization [104, 105]. The boson sector of sound modes is essentially the same as before, but gauge fluctuations are introduced beyond the mean-field analysis and the time sector is modified by the renormalization of the diffusive dynamics of spinons. Nonlocal derivative terms for both space and time should be defined in the momentum and frequency space, where the above expression is just for our formal writing. As discussed before, the boson sector describes the spin-liquid Mott transition, whereas the nonlinear $\sigma$ −model part describes Anderson localization.

We claim that this replica nonlinear $\sigma$ − model approach [107-110] generalizes the Finkelstein's nonlinear $\sigma$ −model approach into the region of Mott transitions. Finkelstein's approach does not incorporate Mott physics associated with strong correlations of electrons. Instead, this approach considers the influences of effective interactions of both singlet and triplet channels in the dynamics of diffusions and Cooperons, considers instabilities of interacting diffusive Fermi-liquids, and reveals the nature of the Anderson MIT. In contrast, the existing nonlinear $\sigma$ −model approach considers instabilities of interacting diffusive spin-liquids, and the nature of the Mott-Anderson MIT. In diffusive Fermi liquids effective interactions of the triplet channel have an important influence on the Anderson MIT of two spatial dimensions [107-110]. In our problem we do not consider triplet-channel interactions, because the positive magneto-electrical resistivity suggests that the magnetic correlations may not be important. However, we speculate that gauge fluctuations, regarded to be singlet-channel interactions, affect the dynamics of Cooperons seriously, thereby suppressing weak-localization corrections and stabilizing the metallic state of spinons. This would be a novel mechanism to produce metallicity near the Mott-Anderson transition. Renormalization group analysis has not been performed yet for this effective field theory.

## 5. Summary

Transition metal perovskites have been the platform for numerous emergent physics that originate from the coupling of the fundamental degrees of freedom such as spin, lattice, charge, and orbital, and also of disorder, which is unavoidable in solids. The physics of 5$d$ perovskite iridates, in particular, has attracted considerable attention after the discovery of the novel $J_{eff}$ = 1/2 Mott insulating state and the evolution of dimensionality controlled MIT in the RP series Sr$_{n+1}$Ir$_n$O$_{3n+1}$, originating due to strong SOC of 5$d$ element which is comparable to its Coulomb correlation or bandwidth. SrIrO$_3$, the end member of the RP series, shows many unusual phenomena including different kinds of MITs. We fabricated epitaxial thin films of various thickness on various substrates to induce MIT in perovskite SrIrO$_3$ by thickness reduction or imposed compressed strain. The MIT driven by thickness reduction occurs due to disorder, but the MIT driven by compressive strain in the films on different substrates is accompanied by peculiar non-Fermi liquid behaviors with an evolving temperature exponent in the electrical resistivity relationship. The latter MIT and associated non-Fermi liquid behaviors are probably due to the delicate interplay between correlation, SOC, and disorder, and thus pose a theoretical challenge to our understanding of non-Fermi liquid physics and MIT.

To reveal the nature of this non-Fermi liquid physics near the MIT of SrIrO$_3$ thin films on various substrates, we first discussed the influence of emergent localized magnetic moments, referred to as UV physics. The observed positive MR in the whole temperature range led us to pursue another direction, referred to as IR physics, in which long-wave length and low-energy fluctuations would be important. Because quantum criticality itself cannot explain the continuous evolution of the non-Fermi liquid physics, we speculate that the interplay between strong correlation of electrons and not-so-weak disorder may contribute to the continuously-varying non-Fermi liquid physics. One possible scenario within IR physics is to consider quantum Griffiths effects with extreme inhomogeneity, in which local fluctuations between metallic and insulating phases, referred to as rare events, and may dominate the non-Fermi liquid physics. To realize this so-called Mott-Anderson-Griffiths scenario, we try to combine spin-liquid Mott physics with Anderson localization that describes a Mott transition from Landau's Fermi-liquid state to a spin-liquid Mott insulating phase. Theoretical understanding of this special class of materials is in its early stages, and many new emergent phenomena in iridates are yet to be explored theoretically or experimentally.

**Acknowledgements**

YHJ acknowledges the support by NRF (Grant No. 2011-0009231). KSK was supported by NRF (Grant No. 2012R1A1B3000550) and also by TJ Park Science Fellowship of the POSCO TJ Park Foundation. YHJ and KSK were jointly supported by CTM at POSTECH (Grant Nos. 2011-0030785 and 2011-0030786).
42

# 6. References

[1] Rao C N R, Raveau B. Transition Metal Oxides: Structure, Properties, and Synthesis of Ceramic Oxides. 2$^{nd}$ Edition. Wiley–VCH; 1998. 392p

[2] Maekawa S, Tohyama T, Barnes S E, Ishihara S, Koshibae W, Khaliullin G. Physics of Transition Metal Oxides. Springer Science & Business Media; 2004. 337p

[3] Tokura Y. Correlated-electron physics in transition-metal oxides. Phys. Today. 2003; **56**: 50-55. (DOI: http://dx.doi.org/10.1063/1.1603080)

[4] Goodenough J B. Perspective on engineering transition-metal oxides. Chem. Mater. 2014; **26**: 820–829. (DOI: 10.1021/cm402063u)

[5] Dagotto E. Complexity in strongly correlated electronic systems. Science. 2005; **309**: 257-262. (DOI: 10.1126/science.1107559)

[6] Johnsson M, Lemmens P. Perovskite and thin films- crystallography and chemistry. J. Phys.: Condens. Matter. 2007; **20**: 264001-6.
(DOI: 10.1088/0953-8984/20/26/264001)

[7] Schlom D G, Chen L-Q, Pan X, Schmehl A, Zurbuchen M A. A thin film approach to engineering functionality into oxides. J. Am. Ceram. Soc, 2008; **91**: 2429-2454. (DOI: 10.1111/j.1551-2916.2008.02556.x)

[8] Martin L W, Chu Y -H, Ramesh R. Advances in the growth and characterization of magnetic, ferroelectric, and multiferroic oxide thin films. Materials Science and Engineering R, 2010; **68**: 89-133.
(DOI:10.1016/j.mser.2010.03.001)

[9] Zubko P, Gariglio S, Gabay M, Ghosez P, Triscone J M. Interface physics in complex oxide heterostructures. Annu. Rev. Condens. Matter Phys., 2011; **2**: 141-65. (DOI: 10.1146/annurev-conmatphys-062910-140445)

[10] Granozio F M, Koster G, Rijnders G. Functional oxide interfaces. MRS Bulletin, 2013; **38**: 1017-1023. (DOI: 10.1557/mrs.2013.282)

[11] Ngai J H, Walker F J, Ahn C H. Correlated oxide physics and electronics. Annu. Rev. Mater. Res., 2014; **44**: 1-17.
(DOI: 10.1146/annurevmatsci 070813-113248)

[12] Sulpizio J A, Ilani S, Irvin P, Levy J. Nanoscale Phenomena in oxide Heterostructures. Annu. Rev. Mater. Res., 2014; **44**: 117-149. (DOI: 10.1146/annurev-matsci-070813-113437)

[13] Chakhalian J, Freeland J W, Millis A J, Panagopoulos, Rondinelli J M. Colloquium: Emergent properties in plane view: Strong correlations at oxide interfaces. Rev. Mod. Phys., 2014; **86**: 1189-1202. (DOI: http://dx.doi.org/10.1103/RevModPhys.86.1189)

[14] Schlom D G, Chen L-Q, Fennie C J, Gopalan V, Muller D A, Pan X, Ramesh R, Uecker R. Elastic strain engineering of ferroic oxides. MRS Bulletin, 2014; **39**: 118-130. (DOI: 10.1557/mrs.2014.1)

[15] Tokura Y, Nagaosa N. Orbital physics in transition-metal oxides. Science, 2000; **288**: 462-468. (DOI: 10.1126/science.288.5465.462)

[16] Cheong S W. Transition metal oxides: The exciting world of orbitals. Nat. Mater., 2007; **6**: 927-928. (DOI:10.1038/nmat2069)

[17] Cao G, De-Long L. Frontiers of 4*d*- and 5*d*-Transition Metal Oxides. 1st Edition. World Scientific; 2013. 328pp

[18] Krempa W W, Chen G, Kim Y B, Balents L. Correlated quantum phenomena in the strong spin-orbit regime. Annu. Rev. Condens. Matter Phys., 2014; **5**: 57-82. (DOI: 10.1146/annurev-conmatphys-020911-125138)

[19] Rau J G, Lee E K –H, Kee H -Y. Spin-Orbit Physics Giving Rise to Novel Phases in Correlated Systems: Iridates and Related Materials. arXiv:1507.06323

[20] Imada M, Fujimori A, Tokura Y. Metal-insulator transitions. Rev. Mod. Phys., 1998; **70**: 1039-1263. (DOI: 10.1103/RevModPhys.70.1039)

[21] Mott N F. Metal-insulator transitions. Taylor & Francis, London; 1990.

[22] Dobrosavljević V, Trivedi N, Valles J M Jr. Conductor insulator quantum phase transitions. Oxford; 2012.

[23] Miranda E, Dobrosavljević V. Disorder-driven non-Fermi liquid behavior of correlated electrons. Rep. Prog. Phys., 2005; **68**: 2337-2408. (DOI:10.1088/0034-4885/68/10/R02)

[24] Edwards P P, Johnston R L, Rao C N R, Tunstall D P, Hensel. The metal-insulator transition: a perspective. Philosophical Transactions A. 1998; **356**: 5-22. (DOI: 10.1098/rsta.1998.0146)

[25] Drude P. Zur Elektronentheorie der metalle. Annalen der Physik. 1900; **306**: 566-613. (DOI: 10.1002/andp.19003060312)

[26] Boer J H, Verwey E J W. Semi-conductors with partially and with completely filled 3*d*-lattice bands. Proceedings of the Phsycial Society. 1937; **49**: 59-71. (DOI:10.1088/0959-5309/49/4S/307)

[27] Mott N F, Peierls R. Discussion of the paper by de Boer and Verwey. Proceedings of the Phsycial Society. 1937; **49**: 72-73. (DOI:10.1088/0959-5309/49/4S/308)

[28] Kotliar G, Vollhardt D. Strongly correlated materials: Insights from dynamical mean-field theory. Physics Today. 2004; **57**: 53-59. (DOI: http://dx.doi.org/10.1063/1.1712502)

[29] Slater J C. Magnetic effects and the Hartree-Fock equation. Phys. Rev., 1951; **82**: 538–541. (DOI: 10.1103/PhysRev.82.538)

[30] Gebhard F. The Mott Metal-Insulator Transition: Models and Methods. Springer; 1997.

[31] Anderson P W. Absence of diffusion in certain random lattices. Phys. Rev., 1958; **109**: 1492-1505. (DOI:10.1103/PhysRev.109.1492)

[32] Lagendijk A, Tiggelen B V, Wiersma D S. Fifty years of Anderson localization. Phys. Today. 2009; **6**2: 24-29. (DOI: http://dx.doi.org/10.1063/1.3206091)

[33] Abrahams E, Anderson P W, Licciardello D C, Ramakrishnan T V. Scaling theory of localization: absence of quantum diffusion in two dimensions. Phys. Rev. Lett., 1979; **42**: 673-676. (DOI:10.1103/PhysRevLett.42.673)

[34] Lee P A, Ramakeishnan T V. Disordered electronic systems. Rev. Mod. Phys., 1985; **57**: 287-337. (DOI: 10.1103/RevModPhys.57.287)

[35] Dolan G J, Osheroff D D. Nonmetallic conduction in thin metal films at low temperatures. Phys. Rev. Lett., 1979; **43**: 721-724. (DOI:10.1103/PhysRevLett.43.721)

[36] Ioffe A F, Regel A R. Semiclassical concept of transport. Prog. Semicond.,1960; **4**: 237

[37] Mott N F. Conduction in non-crystalline systems IX. The minimum metallic conductivity. Phil. Mag., 1972; **26**: 1015-1026. (DOI:10.1080/14786437208226973)

[38] Hussey N E, Takenaka K, Takagi H. Universality of the Mott-Ioffe-Regel limit in metals. Phil. Mag., 2004; **84**: 2847-2864. (DOI:10.1080/14786430410001716944)

[39] Gunnarsson O, Calandra M, Han J E. Colloquium: Saturation of electrical resistivity. Rev. Mod. Phys., 2003; **75**: 1085-1099 (DOI:10.1103/RevModPhys.75.1085)

[40] Licciardello D C, Thouless D J. Constancy of minimum metallic conductivity in two dimensions. Phys. Rev. Lett., 1975; **35**: 1475-1478. (DOI:10.1103/PhysRevLett.35.1475)

[41] Bhalla A S, Guo R, Roy R. The perovskite structure - a review of its role in ceramic science and technology. Mat. Res. Innovations, 2000; **4**: 3-26. (DOI: 10.1007/s100190000062)

[42] Li C, Soh K C K, Wu P. Formability of $ABO_3$ perovskite. J. Alloys Compd., 2004; **372**: 40-48. (DOI: 10.1016/j.jallcom.2003.10.017)

[43] Ruddlesden S N, Popper P. New compounds of the $K_2NiF_4$ type. Acta Crystallogr., 1957; **10**: 538−539. (DOI:10.1107/S0365110X57001929)

[44] Ruddlesden S N, Popper P. The compound $Sr_3Ti_2O_7$ and its structure. Acta Crystallogr., 1958; **11**: 54−55. (DOI:10.1107/S0365110X58000128)

[45] Aurivillius B. Ark. Kemi 1949; **1**: 463−480, 499−512.

[46] Dion M, Ganne M, Tournoux M. Nouvelles familles de phases $M^IM^{II}_2Nb_3O_{10}$ a feuillets "perovskites". Mater. Res. Bull., 1981; **16**: 1429−1435. (DOI:10.1016/0025-5408(81)90063-5)

[47] Jacobson A J, Johnson J W, Lewandowski J T. Interlayer chemistry between thick transition-metal oxide layers: Synthesis and intercalation reactions of $K[Ca_2Na_{n-3}Nb_nO_{3n+1}]$ ($3 \leq n \leq 7$). Inorg. Chem., 1985; **24**: 3727−3729. (DOI: 10.1021/ic00217a006)

[48] Goldschmidt V M. Die gesetze der krystallochemie. Naturwissenschaften, 1926; **14**: 477-485. (DOI: 10.1007/BF01507527)

[49] Cullity B D. Introduction to Magnetic Materials. Addison-Wesley; 1972.

[50] Moon S J, Jin H, Kim K W, Choi W S, Lee Y S, Yu J, Cao G, Sumi A, Funajubo H, Bernhard C, Noh T W. Dimensionality-controlled insulator-metal transition and correlated metallic state in 5d transition metal oxides $Sr_{n+1}Ir_nO_{3n+1}$ ($n$ = 1, 2, and ∞). Phys. Rev. Lett., 2008; **101**: 226402-4. (DOI:10.1103/PhysRevLett.101.226402)

[51] Crawford M K, Subramanian M A, Harlow R L, Fernandez-Baca, Wang Z R, Johnston D C. Structural and magnetic studies of $Sr_2IrO_4$. Phys. Rev. B., 1994; **49**: 9198-9201. (DOI:10.1103/PhysRevB.49.9198)

[52] Cao G, Bolivar J, McCall S, Crow J E, Guertin R P. Weak ferromagnetism, metal-to-nonmetal transition, and negative differential resistivity in single-crystal $Sr_2IrO_4$. Phys. Rev. B (R). 1998; **57**: 11039-11042. (DOI:http://dx.doi.org/10.1103/PhysRevB.57.R11039)

[53] Kim B J, Jin H, Moon S J, Kim J –Y, Park B –G, Leem C S, Yu J, Noh T W, Kim C, Oh S -J, Park J –H, Durairaj V, Rotenberg E. Novel $J_{eff}$=1/2 Mott state induced by relativistic spin-orbit coupling in $Sr_2IrO_4$. Phys. Rev. Lett., 2008; **101**: 076402-4. (DOI:10.1103/PhysRevLett.101.076402)

[54] Kim B J, Ohsumi H, Komesu T, Sakai S, Morita T, Takagi H, Arima T. Phase-sensitive observation of a spin-orbital mott state in $Sr_2IrO_4$. Science. 2009; **323**: 1329-1332. (DOI: 10.1126/science.1167106)

[55] Wojek B M, Bernsten M H, Boseggia S, Boothroyd A T, Prabhakaran D, Mcmorrow D F, Ronnow H M, Chang J, Tjernberg O. The $J_{eff}$=1/2 insulator $Sr_3Ir_2O_7$ studied by means of angle-resolved photoemission spectroscopy. J. Phys. Condens. Matter. 2012; **24**: 415602-4. (DOI:10.1088/0953-8984/24/41/415602)

[56] Nichols J, Terzic J, Bittle E G, Korneta O B, Long L E D, Brill J W, Cao G, Seo S S A. Tuning electronic structure via epitaxial strain in $Sr_2IrO_4$ thin films. Appl. Phys. Lett., 2013; **102**: 141908-4. (DOI:http://dx.doi.org/10.1063/1.4801877)

[57] Zocco D A, Hamlin J J, White B D, Kim B J, Jeffries J R, Weir S T, Vohra Y K, Allen J W, Maple M B. Persistent non-metallic behavior in $Sr_2IrO_4$ and $Sr_3Ir_2O_7$ at high pressures. J. Phys. Condens. Matter. 2014; **26**: 255603-4. (DOI:10.1088/0953-8984/26/25/255603)

[58] Domingo N, López-Mir L, Paradinas M, Holy V, Zelezny J, D, Suresha S J, Liu J, Rayan-Serrao C, Ramesh R, Ocal C, Martí X, Catalan G. Giant reversible nanoscale piezoresistance at room temperature in $Sr_2IrO_4$ thin films. Nanoscale. 2015; **7**: 3453-3459. (DOI: 10.1039/C4NR06954D)

[59] Wang C, Seinige H, Cao G, Zhou J -S, Goodenough J B, Tsoi M. Electrically tunable band gap in antiferromagentic Mott insulator $Sr_2IrO_4$. arXiv:1502.07982

[60] Kim Y K, Krupin O, Denlinger J D, Bostwick A, Rotenberg E, Zhao Q, Mitchell J F, Allen J W, Kim B J. Science. Fermi arcs in a doped pseudospin-1/2 Heisenberg antiferromagnet. 2014; **345**: 187-190. (DOI:10.1126/science.1251151)

[61] Lupascu A, Clancy J P, Gretarsson H, Nie Z, Nichols J, Terzic J, Cao G, Seo S S A, Islam Z, Upton M H, Kim J, Casa D, Gog T, Said A H, Katukuri V M, Stoll H, Hozoi L, van der Brink J, Kim Y-J. Tuning magnetic coupling in $Sr_2IrO_4$ thin films with epitaxial Strain. Phys. Rev. Lett., 2014; **112**: 147201-5. (DOI: 10.1103/PhysRevLett.112.147201)

[62] Ge M, Qi T F, Korneta O B, De Long D E, Schlottmann P, Crummett W P, Cao G. Lattice-driven magnetoresistivity and metal-insulator transition in single-layered iridates. Phys. Rev. B (R). 2011; **84**: 100402-5. (DOI:10.1103/PhysRevB.84.100402)

[63] Fujiyama S, Ohsumi H, Ohashi K, Hirai D, Kim B J, Arima T, Takata M, Takagi H. Spin and orbital contributions to magnetically ordered moments in 5d layered perovskite $Sr_2IrO_4$. Phys. Rev. Lett., 2014; **112**: 016405-5. (DOI:10.1103/PhysRevLett.112.016405)

[64] Qi T F, Korneta O B, Li L, Butrouna K, Cao V S, Wan X, Schlottmann P, Kaul R K, Cao G. Spin-orbit tuned metal-insulator transitions in single-crystal $Sr_2Ir_{1-x}Rh_xO_4$ ($0 \leq x \leq 1$). Phys. Rev. B. 2012; **86**: 125105-6. (DOI:10.1103/PhysRevB.86.125105)

[65] Wang C, Seinige H, Cao G, Zhou J -S, Goodenough J -B, Tsoi M. Anisotropic magnetoresistance in antiferromagnetic $Sr_2IrO_4$. Phys. Rev. X. 2014; **4**:041034-5. (DOI:10.1103/PhysRevX.4.041034)

[66] Kim J, Daghofer M, Said A H, Gog T, van den Brink J, Khaliullin G , Kim B J. Excitonic quasiparticles in a spin-orbit Mott insulator. Nat. Commun., 2014; **5**: 4453-6. (DOI:10.1038/ncomms5453)

[67] Subramanian M A, Crawford M K, Harlow R L. Single crystal structure determination of double layered strontium iridium oxide [$Sr_3Ir_2O_7$]. Mat. Res. Bull., 1994; **29**: 645-650. (DOI: 10.1016/0025-5408(94)90120-1)

[68] Li L, Kong P P, Qi T F, Jin C Q, Yuan S J, DeLong L E, Schlottmann P, Cao G. Tuning the $J_{eff}$=1/2 insualting state via electron doping and pressure in the double-layered iridtaes $Sr_3Ir_2O_7$. Phys. Rev. B., 2013; **87**: 235127-6. (DOI:10.1103/PhysRevB.87.235127)

[69] Cao G, Xin Y, Alexander C S, Crow J E, Schlottmann P, Crawford M K, Hatlow R L, Marshall W. Anomalous magnetic and transport behavior in the magnetic insulator $Sr_3Ir_2O_7$. Phys. Rev. B., 2002; **66**: 214412-7. (DOI:10.1103/PhysRevB.66.214412)

[70] Kim J W, Choi Y, Kim J, Mitchell J F, Jackeli G, Daghofer M, van der Brink J, Khaliullin G, Kim B J. Dimensionality driven spin-flop transition in layered iridates. Phys. Rev. Lett. 2012; **109**: 037204-5. (DOI:10.1103/PhysRevLett.109.037204)

[71] Li L, Kong P P, Qi T F, Jin C Q, Yuan S J, DeLong L E, Schlottmann P, Cao G. Tuning *J*eff=1/2 insualting state via electron doping and pressure in double-layered iridate $Sr_3Ir_2O_7$. Phys. Rev. B., 2013; **87**: 235127-6. (DOI: 10.1103/PhysRevB.87.235127)

[72] Kim J, Said A H, Casa D, Upton M H, Gog T, Daghofer M, Jackeli G, van den Brink J, Khaliullin G, Kim B J. Large spin-wave energy gap in the bilayer iridate $Sr_3Ir_2O_7$: Evidence for the enhanced dipolar interactions near the Mott metal-insulator transitions. Phys. Rev. Lett. 2012; **109**: 157402-5. (DOI:10.1103/PhysRevLett.109.157402)

[73] Okada Y, Walkup D, Lin H, Dhital C, Chang T -R, Khadka S, Zhou W, Jeng H -T, Bansil A, Wang Z, Wilson S, Madhavan V. Imaging the evolution of metallic states in a correlated iridates. Nat. Mater. 2013; **12**: 707-713. (DOI:10.1038/nmat3653)

[74] Park H J, Sohn C H, Jeong D W, Cao G, Kim K W, Moon S J, Jin H, Cho D -Y, Noh T W. Phonon-assisted optical excitation in the narrow bandgap Mott insulator $Sr_3Ir_2O_7$. Phys. Rev. B. 2014; **89**: 155115-6. (DOI:10.1103/PhysRevB.89.155115)

[75] Longo J M, Kafalas J A, Arnott R J. Structure and properties of the high and low pressure forms of $SrIrO_3$. Journal of Solid State Chemistry. 1971; **3**: 174-179. (DOI:10.1016/0022-4596(71)90022-3)

[76] Cao G, Durairaj V, Chikara S, DeLong L E, Parkin S, Schlottmann P. Non-Fermi liquid behavior in nearly ferromagnetic $SrIrO_3$ single crystals. Phys. Rev. B. (R), 2007; **76**: 100402-4. (DOI: 10.1103/PhysRevB.76.100402)

[77] Zhao J G, Yang L X, Yu Y, Li F Y, Yu R C, Fang Z, Chen L C, Jin C Q. High-pressure synthesis of orthorhombic $SrIrO_3$ perovskite and its positive magnetoresistance. J. Appl. Phys. 2008; **103**: 103706-5. (DOI:10.1063/1.2908879)

[78] Blanchard P E R, Reynolds E, Kennedy B J, Kimpton J A, Avdev M, Belik A A. Anomalous thermal expansion in orthorhombic peovskite $SrIrO_3$: Interplay between spin-orbit coupling and the crystal lattice. Phys. Rev. B. 2014; **89**: 214106-8. (DOI:10.1103/PhysRevB.89.214106)

[79] Nie Y F, King P D C, Kim C H, Uchida M, Wei H I, Faeth B D, Ruf J P, Ruff J P C, Xie L, Pan X, Fennie C J, Schlom D G, Shen K M. Interplay of spin-orbit interactions, dimensionality, and octahedral rotations in semimetallic $SrIrO_3$. Phys. Rev. Lett. 2015; **114**: 016401-6. (DOI:10.1103/PhysRevLett.114.016401)

[80] Carter J -M, Shankar V V, Zeb M A, Kee H –Y. Semimetal and topological insulator in perovsakite iridates. Phys. Rev. B. 2012; **85**: 115105-6. (DOI:10.1103/PhysRevB.85.115105)

[81] Chen Y, Lu Y –M, Kee H –Y. Toplogical crystalline metal in orthorhombic perovskite iridates. Nat. Commun. 2015; **6**: 6593-7. (DOI:10.1038/ncomms7593)

[82] Zeb M A, Kee H –Y. Interplay between spin-orbit coupling and Hubbard interaction in $SrIrO_3$ and related Pbnm perovskites. Phys. Rev. B. 2012; **86**: 085149-7. (DOI:10.1103/PhysRevB.86.085149)

[83] Biswas A, Kim K –S, Jeong Y H. Metal insulator transitions in perovskite SrIrO$_3$ thin films. J. Appl. Phys. 2014; **116**: 213704-10. (DOI: http://dx.doi.org/10.1063/1.4903314)

[84] Wu F X, Zhou J, Zhang L Y, Chen Y B, Zhang S –T, Gu Z –B, Yao S – H, Chen Y –F, Metal-insulator transition in SrIrO$_3$ with strong spin-orbit interaction. J. Phys.: Condens. Matter. 2013; **25**: 125604-8. (DOI:10.1088/0953-8984/25/12/125604)

[85] Yoshimatsu K, Okabe T, Kumihashira H, Okamoto S, Aizaki S, Fujimori A, Oshima M. Dimensional-crossover-driven metal-insulator transition in SrVO$_3$ ultrathin Films. Phys. Rev. Lett. 2010; **104**: 147601-4. (DOI:10.1103/PhysRevLett.104.147601)

[86] Gruenewald J H, Nichols J, Terzic J, Cao G, Brill J W, Seo S S A. Compressive strain-induced metal-insulator transition in orthorhombic SrIrO$_3$ thin films. J. Mater. Res. 2014; **29**: 2491-2496. (DOI:10.1557/jmr.2014.288)

[87] Zhang L, Liang Q, Xiong Y, Zhang B, Gao L, Handong Li H, Chen Y B, Zhou J, Zhang S -T, Gu Z -B, Yao S, Wang Z, Lin Y, Chen Y –F. Tunable semimetallic state in compressive-strained SrIrO$_3$ films revealed by transport behaviors. Phys. Rev. B. 2015; **91**: 035110-9. (DOI:10.1103/PhysRevB.91.035110)

[88] Georges A, Kotliar G, Krauth W, Rozenberg M J. Dynamical mean-field theory of strongly correlated fermion systems and the limit of infinite dimension. Rev. Mod. Phys. 1996; **68**: 13-125. (DOI:http://dx.doi.org/10.1103/RevModPhys.68.13)

[89] Vucicevic J, Terletska H, Tanaskovic D, Dobrosavljevic V. Finite-temperature crossover and the quantum Widom line near the Mott transition. Phys Rev B. 2013; **88**: 075143-12. (DOI: 10.1103/PhysRevB.88.075143)

[90] Deng X, Mravlje J, Žitko R, Ferrero M, Kotliar G, Georges A. How bad metal turn good: Spectroscopic signature of resilient quasiparticles. Phys. Rev. Lett. 2013; **110**: 086401-5. (DOI:10.1103/PhysRevLett.110.086401)

[91] Löhneysen H, Rosch A, Vojta M, Wölfle P. Fermi-liquid instabilities at magnetic quantum phase transition. Rev. Mod. Phys. 2007; **79**: 1015-1075. **(**DOI:10.1103/RevModPhys.79.1015**)**

[92] Harris A B. Nature of the "Griffiths" singularity in dilute magnets. Phys. Rev. B. 1975; **12**: 203-207. (DOI: 10.1103/PhysRevB.12.203)

[93] Vojta T. Phase and phase transitions in disordered quantum systems. AIP Conf. Proc. 2013; 1550: 188-247. (DOI: 10.1063/1.4818403)

[94] Bhatt R N, Lee P A. Scaling studies of highly disordered spin-1/2 antiferromagnetic systems. Phys. Rev. Lett. 1982; **48**: 344-347. (DOI:10.1103/PhysRevLett.48.344)

[95] Griffiths R B. Nonanalytic behavior above the crtitcal point in a random ising ferromagnet. Phys. Rev. Lett. 1969; **23**: 17-19. (DOI:10.1103/PhysRevLett.23.17)

[96] Byczuk K. Metal-insulator transitions in the Falicov-Kimball model with disorder. Phys. Rev. B. 2005; 71: 205105-8. (DOI:10.1103/PhysRevB.71.205105)

[97] Ruderman M A, Kittel C. Indirect exchange coupling of nuclear magnetic moments by conduction electrons. Phys. Rev. 1954; **96**: 99-102. (DOI:http://dx.doi.org/10.1103/PhysRev.96.99)

[98] Kasuya T. A theory of metallic ferro-and antiferromagentism on Zener's model. Prog. Theor. Phys. 1956; **16**: 45-57. (DOI: 10.1143/PTP.16.45)

[99] Yosida K. Magnetic Properties of Cu-Mn Alloys. Phys. Rev.1957; **106**: 893-898. (DOI: 10.1103/PhysRev.106.893)

[100] Lee P A, Nagaosa N, Wen X -G. Doping a Mott insulator: Physics of high-temperature superconductivity. Rev. Mod. Phys. 2006; **7**: 17-85. (DOI:http://dx.doi.org/10.1103/RevModPhys.78.17)

[101] Florens S, Georges A. Slave-rotor mean-field theories of strongly correlated systems and the Mott transition in finite dimensions. Phys. Rev. B 2004; **70**: 035114-15. (DOI: 10.1103/PhysRevB.70.035114)

[102] Cho D, Cheon S, Kim K -S, Lee S -H, Cho Y -H, Cheong S -W, Yeom H W. Nanoscale manipulation of the Mott insulating state coupled to charge order in 1T-$TaS_2$. arXiv:1505.00690

[103] Kim K –S. Role of disorder in the Mott-Hubbard transition. Phys. Rev. B 2006; 73: 235115-10. (DOI: 10.1103/PhysRevB.73.235115)

[104]	Wegner F. The mobility edge problem: Continuous symmetry and a conjecture. Z. Phys. B. 1979; 35: 207-210. (DOI: 10.1007/BF01319839)

[105]	Sch¨afer L, Wegner F. Disordered system with n orbitals per Site: Lagrange formulation, hyperbolic symmetry, and goldstone modes. Z. Phys. B 1980; 3: 113-126. (DOI: 10.1007/BF01598751)

[106]	Kim K – M, Kim K –S. in preparation

[107]	Finkel'stein A M. Influence of coulomb interaction on the properties of disordered metals. Sov. Phys. JETP 1983; **57**: 97-108

[108]	Finkel'stein. A M. Electron liquid in disordered conductors. Sov. Sci. Rev. London; 1990. Vol. 14. 1-101p.

[109]	Punnoose A, Finkelstein A M. Metal-insulator transition in disordered two-dimensional electron systems. Science 2005; **310**: 289-291. (DOI:10.1126/science.1115660)

[110]	Anissimova S, Kravchenko S V, Punnoose A, Finkel'stein A M, Klapwijk T M. Flow diagram of the metal–insulator transition in two dimensions. Nat. Phys. 2007; **3**: 707-710. (DOI: 10.1038/nphys685)